\documentclass[superscriptaddress,aps,prd,nofootinbib,twocolumn,showpacs, floatfix, 10pt]{revtex4-2}

\usepackage{float}
\usepackage{amsfonts}
\usepackage{amsmath}
\usepackage{amssymb}
\usepackage{graphicx}
\usepackage{mathrsfs}
\usepackage{hhline,color}
\usepackage{pifont,ulem}
\usepackage{lmodern}
\usepackage{epsfig}
\usepackage{enumitem} 
\usepackage{ulem}
\usepackage[dvipsnames]{xcolor}

\usepackage{subcaption}

\usepackage{ragged2e} 
\DeclareCaptionJustification{justified}{\justifying}
\usepackage{comment}

\usepackage[utf8]{inputenc}
\usepackage{multirow}
\usepackage{lipsum}

\usepackage{braket}

\usepackage{hyperref}

\newcommand{\psibar}{\bar{\psi}}
\newcommand{\F}{{\cal F}}
\newcommand{\Phibar}{\bar \Phi}
\newcommand{\muR}{\mu^*}
\newcommand{\diff}{{\rm d}}

\def\beq{\begin{equation}}
\def\eeq{\end{equation}}
\def\beqa{\begin{eqnarray}}
\def\eeqa{\end{eqnarray}}
\def\ban{\begin{eqnarray*}}
\def\ean{\end{eqnarray*}}
\def\bi{\begin{itemize}}
\def\ei{\end{itemize}}

\begin{document}

\title{Isentropic thermodynamics across the hadron-quark mixed phase in a 
two-phase model with a PNJL quark description}

\author{Eduardo L. G. Salgado}
\email{uc2020216840@student.uc.pt}
\affiliation{CFisUC, Department of Physics,
University of Coimbra, P-3004 - 516  Coimbra, Portugal}
\author{Pedro Costa}
\email{pcosta@uc.pt}
\affiliation{CFisUC, Department of Physics,
University of Coimbra, P-3004 - 516  Coimbra, Portugal}
\author{Constan\c ca Provid\^encia}
\email{cp@fis.uc.pt}
\affiliation{CFisUC, Department of Physics,
University of Coimbra, P-3004 - 516  Coimbra, Portugal}

\date{\today}

\begin{abstract}
We study the hadron-quark mixed phase within a two-phase model for symmetric and asymmetric matter. For the quark sector we employ the (2+1) Polyakov-extended Nambu–Jona-Lasinio model (PNJL) with vector interactions. We investigate how the hadronic equation of state affects the phase diagram and the thermodynamic properties inside the mixed phase. The behavior of isentropic trajectories in the mixed phase depends on the fixed entropy per baryon ($s/\rho_B$), with trajectories near the critical end point (CEP) exhibiting a pronounced cooling pattern, while isentropic trajectories with low entropy per baryon undergo pronounced  heating as the baryonic density increases. The adiabatic squared speed of sound displays characteristic peak and dip structures that depend on $s/\rho_B$.  The polytropic index along isentropic and isothermal trajectories, including in the vicinity of the CEP are also investigated. The effects of vector interactions and isospin asymmetry on thermodynamic observables likewise depend on the chosen $s/\rho_B$ value.  Finally, we discuss the population of hyperons along isentropic trajectories and their influence on the phase diagram. The main effect of hyperons is to shift the onset of deconfinement to larger densities and decrease the density extension of the mixed phase.
\end{abstract}

\maketitle

\section{Introduction}

{Under extreme conditions of temperature and baryon density, such as those reached in relativistic heavy-ion collisions or in the interior of compact stars, confinement is expected to give way to deconfinement, while chiral symmetry becomes (partially) restored. Under these conditions, strongly interacting matter may transition to a deconfined quark phase or form a quark-gluon plasma (QGP). Mapping the phase structure of QCD matter and identifying observable signatures of the hadron–quark transition are therefore central objectives of contemporary nuclear and particle physics.

Lattice QCD calculations indicate that, at finite temperature and vanishing baryon chemical potential, the transition from hadronic matter to the QGP is not a genuine phase transition but rather a smooth crossover~\cite{Aoki:2006we, Borsanyi:2010bp, HotQCD:2019xnw, Bazavov:2011nk, Kotov:2021rah, Cuteri:2021ikv}. At finite baryon chemical potential, however, first-principles lattice calculations are limited by the fermion sign problem, leaving the structure of the QCD phase diagram in this region largely uncertain. A widely discussed scenario predicts the existence of a first-order phase transition at sufficiently large baryon chemical potentials, terminating at a critical end point (CEP) where the transition becomes second order. Support for such a structure has been obtained in various effective approaches, including the Nambu–Jona-Lasinio (NJL) model and the quark–meson (QM) model, without and with Polyakov loop, as well as in functional frameworks such as Dyson–Schwinger equations (DSE) and the functional renormalization group (FRG)~\cite{Costa:2008gr,Ferreira:2017wtx,Cao:2025zvh,CamaraPereira:2020xla, Bzdak:2019pkr, Du:2024wjm, Eichmann:2015kfa, Fu:2019hdw}.

Experimentally, locating the CEP and determining the nature of the hadron–quark phase transition are primary goals of ongoing and planned heavy-ion programs, including those at RHIC, the SPS and LHC at CERN, and the upcoming FAIR facility at GSI. In particular, data from the second phase of the RHIC Beam Energy Scan (BES-II) program have achieved improved precision in measurements of net-proton cumulants and revealed possible nonmonotonic trends that may be indicative of critical phenomena~\cite{STAR:2025zdq}. In this context, theoretical studies of the evolution along isentropic trajectories and the associated collective flow dynamics are essential for interpreting experimental observations.}

A first-order hadron–quark phase transition implies the existence of a mixed-phase region in which hadronic and quark degrees of freedom coexist. A common and physically transparent framework to describe this regime is the two-phase construction, in which hadronic and quark matter are modeled separately and matched through the appropriate phase-equilibrium conditions. This approach has long been employed in the study of neutron-star matter under weak equilibrium~\cite{Glendenning:1992vb, Glendenning:1997ak, Dexheimer:2009hi, Shao:2011nu}, and has also been applied to heavy-ion collision scenarios, particularly for isospin-asymmetric matter~\cite{DiToro:2006bkw, DiToro:2008zm, DiToro:2009ig, Cavagnoli:2010yb, Shao:2011ij, Shao:2011fk}. Strangeness and vector interactions also play a crucial role in determining the properties of  QCD matter. In the quark sector, repulsive vector couplings tend to shift the phase transition toward higher baryon densities and lower temperatures and may significantly modify the location of the CEP or even eliminate it~\cite{Fukushima:2008wg, Costa:2010zw,Costa:2016vbb,Denke:2013gha, Ferreira:2018sun}. In the hadronic sector, isoscalar and isovector interactions govern the stiffness of the equation of state (EoS) and control the system’s response to isospin asymmetry. Moreover, as temperature and baryon density increase, the appearance of hyperons becomes energetically favorable, signaling the transition from purely nucleonic to hyperonic matter. The emergence of these additional degrees of freedom generally softens the hadronic EoS and influences the onset of quark matter, the characteristics of the hadron–quark phase transition, and the thermodynamic properties of the system as a whole.

An important feature of the fireball produced in HICs is its approximately adiabatic expansion. As a consequence, the system evolves along trajectories of nearly constant entropy per baryon in the phase diagram. The behavior of thermodynamic quantities along these isentropic trajectories provides valuable insight into the nature of the phase transition and its possible experimental signatures. The speed of sound is one of the key quantities of interest along isentropic trajectories, as it reflects the stiffness of the EoS and directly influences the hydrodynamic evolution of the fireball. Recently, in Ref.~\cite{Sorensen:2021zme}, the authors estimated the speed of sound and its logarithmic derivative with respect to baryon density, and explored their connection with cumulants of the baryon-number distribution in matter created in HICs, with the aim of identifying signatures of the QCD CEP. Furthermore, Ref.~\cite{Ferreira:2018sun} showed that the existence of a CEP cannot be established solely on the basis of nonmonotonic behavior in net-baryon number susceptibilities, since such behavior may also arise in the absence of a CEP. This observation motivates the investigation of additional observables that may serve as complementary indicators of the hadron–quark phase transition. The polytropic index is another useful quantity that can quantitatively distinguish between hadronic and quark matter, and has been widely employed in studies of neutron-star matter~\cite{Annala:2017llu, Liu:2022mje, Liu:2023gmq, Liu:2023kzy}.

In this work, the hadron–quark phase transition is investigated within a two-phase framework, considering both isospin-symmetric and isospin-asymmetric matter. The hadronic phase is described using {relativistic mean-field models~\cite{Dutra:2014qga}}, extended to include the full baryon octet. The deconfined phase is modeled using the $(2+1)$-flavor Polyakov–Nambu–Jona-Lasinio (PNJL) model~\cite{Fukushima:2003fw, Fu:2007xc, Ciminale:2007sr, Fukushima:2008wg, Abuki:2008nm, Costa:2010zw}, which incorporates chiral symmetry breaking and an effective description of deconfinement through the Polyakov loop. Repulsive vector interactions among quarks are also included. Our analysis focuses primarily on the behavior of strongly interacting matter along isentropic trajectories, both across the hadron–quark coexistence region and within the pure phases. Particular attention is given to the influence of vector interactions, isospin asymmetry, and hyperonic degrees of freedom on the location and extent of the mixed phase. Along these trajectories, we compute the speed of sound and the polytropic index, and examine their behavior in the vicinity of the hadron–quark phase transition boundary and the CEP.

\section{Model and formalism} 
\label{sec:model}

In this section, we present the EoSs used for the hadronic and quark phases within the two-phase model employed throughout this work. Quark vector interaction and the possible appearance of hyperons in the hadronic phase are included.  In the mixed phase the pure hadronic and quark phases are connected by the Gibbs conditions (thermal, chemical, and mechanical equilibrium) under conservation of baryon number and isospin in the strong-interacting process. 

\subsection{Quark phase}
Quark matter is described by the Lagrangian of the  $(2+1)$-flavor PNJL model,
\begin{eqnarray}\label{Pnjl}
    {\cal L}_{\rm PNJL} 
    &=& {\psibar} \left(i\gamma_\mu D^{\mu}-{ m} + { \mu \gamma_0}     \right ) \psi
        + {\cal L}_\text{sym}
        +{\cal L}_\text{det}
        +{\cal L}_\text{vec} 
        \notag\\
        &-&\mathcal{U}\left(\Phi,\bar\Phi;T\right),
\end{eqnarray}
where  $\psi = (u,d,s)^T$  and 
${ m}= {\rm diag}_f (m_u,m_d,m_s)$ is the corresponding (current) mass  matrix in flavor-space. Finite-density effects are introduced by the chemical
potential matrix ${\mu} = {\rm diag}_f (\mu_u, \mu_d, \mu_s)$. 
The scalar-pseudoscalar interaction reads
\begin{equation}
	{\cal L}_\text{sym}= G_S \sum_{a=0}^8 \left [({\bar \psi} \lambda_ a \psi)^2 + 
	({\bar \psi} i\gamma_5 \lambda_a \psi)^2 \right ],
\end{equation}
which  spontaneously breaks chiral symmetry in the vacuum. The Kobayashi-Maskawa-'t Hooft interaction,
\begin{equation}
	{\cal L}_\text{det}=-G_D\left({\rm det} \left [{\bar \psi}(1+\gamma_5)\psi \right] + 
	{\rm det}\left [{\bar \psi}(1-\gamma_5)\psi\right] \right ),
\end{equation} 
generates the six-fermion flavor-mixing term that explicitly breaks the $U_A(1)$ symmetry and correctly reproduces the observed hadron spectrum~\cite{Klevansky:1992qe,Hatsuda:1994pi}.

We include a vector interaction of the form~\cite{Mishustin:2000ss}
\begin{equation} 
{\cal L}_\text{vec} = - G_V \sum_{a=0}^8  
\left[({\bar \psi} \gamma^\mu \lambda_a \psi)^2 + 
 ({\bar \psi} \gamma^\mu \gamma_5 \lambda_a \psi)^2 \right]. 
\label{p1} 
\end{equation}
Rather than attempting to fix $G_V$ from vacuum vector-meson properties, we study its in-medium effect by varying the ratio $\zeta = G_V/G_S$. Vector interactions are known to be important in the study of neutron stars~\cite{Bonanno:2011ch, CamaraPereira:2016chj}
or the QCD phase diagram~\cite{Ferreira:2018sun}, and may have a non-trivial density dependence. 

The quark fields are minimally coupled to a static background gluon field $A^0_4$,
via the covariant derivative $D_\mu = \partial_\mu -A^0_4 \delta_\mu^0$. The Polyakov loop $\Phi$ acts as an order parameter for the center symmetry in the pure-gauge limit: in the confined phase $\Phi \to 0$, while $\Phi \to 1$ in the deconfined phase. In this model, the effective potential $\mathcal{U}(\Phi, \bar\Phi; T)$ is responsible for describing the statistical confinement-deconfinement transition, with the breaking of $Z(N_c)$ symmetry. 
We choose to adopt the effective Polyakov potential of Refs.~\cite{Roessner:2006xn},
\begin{align} \label{Ueff}
	\frac{\mathcal{U}\left(\Phi,\bar\Phi;T\right)}{T^4}&= -\frac{1}{2}{a\left(T\right)}\bar\Phi \Phi
	\notag \\
    &+ b(T)\mbox{ln}\left[1-6\bar\Phi\Phi +4(\bar\Phi^3+ \Phi^3) -3(\bar\Phi \Phi)^2\right],
\end{align}
where 
$a\left(T\right)=a_0+a_1\left(\frac{T_0}{T}\right)+a_2\left(\frac{T_0}{T}\right)^2$, 
$b(T)=b_3\left(\frac{T_0}{T}\right)^3$. 
Its parametrization values are $a_0 = 3.51$, $a_1 = -2.47$, $a_2 = 15.2$, 
and $b_3 = -1.75$~\cite{Roessner:2006xn}, while the critical temperature is set 
to $T_0=210$ MeV.
The divergent ultraviolet sea quark integrals are regularized by a sharp cutoff 
$\Lambda$ in three-momentum space. For the NJL model parametrization, we consider:
$\Lambda = 630$ MeV, $m_u= m_d=5.5$ MeV, $m_s=135.7$ MeV, 
$G_s \Lambda^2= 1.781$, and $G_D \Lambda^5=9.29$~\cite{CamaraPereira:2016chj}.
The quark condensates and the Polyakov loop value are determined in the mean-field approximation (see App.~\ref{sec:app_formalism})

\subsection{Hadron phase}
We describe hadronic matter within a relativistic mean-field (RMF) framework. To assess the impact of hyperons we employ two parameterizations: one for nucleonic matter only and one including the full baryon octet. In these models nucleons are coupled to the neutral scalar $\sigma$,  the isoscalar-vector $\omega^\mu$,  and  the isovector-vector $\boldsymbol{\rho}^\mu$. When hyperons are included we also add the isoscalar–vector $\phi^\mu$, which mediates repulsive interactions among hyperons and tends to stiffen the equation of state.

The Lagrangian density of the theory is given by
\begin{equation}\label{hadr_general}
    {\cal L} = \sum_b {\cal L}_b + {\cal L}_m,
\end{equation}
where ${\cal L}_b$ and ${\cal L}_m$ denote the baryon and meson contributions, respectively. For nucleonic matter (neutrons $n$ and protons $p$) we  use the NL3$\omega\rho$ model~\cite{Lalazissis:1996rd, Carriere:2002bx}. The baryonic term reads
\begin{eqnarray}\label{lb_nl3wr}
    {\cal L}_b
    &=&
    \psibar_b\left(
                i\gamma_\mu\partial^\mu  - m_b + g_{\sigma b}\sigma -g_{\omega b}\gamma_\mu\omega^\mu\right.
                - \left. g_{\rho b} \gamma_\mu \boldsymbol{\tau}_b\cdot \boldsymbol{\rho}^\mu\right.
            \notag\\ 
                &+&\left.\mu_b\gamma_0 \right)\psi_b,
\end{eqnarray}
with $b \in \{n,p\}$, $\psi_b$ the baryon Dirac field and $\boldsymbol{\tau_b}$ the isospin operator. The mesonic term is
\begin{eqnarray}\label{lm_nl3wr}
    {\cal L}_m
    &&=
        \frac{1}{2}\partial_\mu\sigma\partial^\mu\sigma - \frac{1}{2}m_\sigma^2 \sigma^2 - \frac{\kappa}{3!}(g_{\sigma N}\sigma)^3 - \frac{\lambda}{4!}(g_{\sigma N}\sigma)^4
    \notag\\
         &&- \frac{1}{4}\omega_{\mu\nu}\omega^{\mu\nu} + \frac{1}{2} m^2_{\omega}\omega_{\mu}\omega^{\mu} + \frac{\xi}{4!}g_{\omega N}^4 (\omega_\mu\omega^\mu)^2
    \notag\\
        &&-\frac{1}{4}\boldsymbol{\rho}_{\mu\nu}\boldsymbol{\rho}^{\mu\nu} + \frac{1}{2}m^2_\rho\boldsymbol{\rho}_\mu\boldsymbol{\rho}^\mu
        \notag\\ 
        && + \Lambda_\omega g^2_{\rho N}g_{\omega N}^2\boldsymbol{\rho}_\mu\boldsymbol{\rho}^\mu\omega_\mu\omega^\mu.
\end{eqnarray}
where the field strength tensors are defined by $\omega_{\mu\nu} = \partial_\mu\omega_\nu - \partial_\nu\omega_\mu$ and $\boldsymbol{\rho}_{\mu\nu} = \partial_\mu\boldsymbol{\rho}_\nu - \partial_\nu\boldsymbol{\rho}_\mu$. The couplings $\kappa$ and $\lambda$ define, respectively, the strength of the cubic and the quartic self-interacting $\sigma$-meson terms and
the coupling $\xi$ denotes the strength of $\omega$-meson quartic term \cite{Horowitz:2000xj}.

To include hyperons we extend the baryon sum to the full spin-$1/2$ octet,
\begin{equation}
\begin{split}
    b \in \{n,p\} &\to \{n,p,\Lambda, \Sigma^\pm, \Sigma^0, \Xi^0, \Xi^-\},
\end{split}
\end{equation}
and introduce the $\phi$-meson, which couples to the hyperons. The baryonic Lagrangian is modified as
\begin{equation}
     {\cal L}_b \to {\cal L}_b - \psibar_b\, g_{\phi b}\gamma_\mu\phi^\mu \,\psi_b,
\end{equation}
while for the mesonic Lagrangian we add
\begin{equation}
    {\cal L}_m \to  {\cal L}_m - \frac{1}{4}\phi_{\mu\nu}\phi^{\mu\nu} + \frac{1}{2}m^2_\phi\phi_\mu\phi^\mu,
\end{equation}
where $\phi_{\mu\nu} = \partial_\mu\phi_\nu - \partial_\nu\phi_\mu$ and the sum in Eq.~\eqref{hadr_general} runs over the baryonic octet, the nucleons together with the hyperon states $\Lambda$, $\Sigma^+$, $\Sigma^0$, $\Sigma^-$, $\Xi^0$, $\Xi^-$. The $\Delta$ baryon is ignored as it is too heavy to be relevant in the regime considered. When hyperons are included we employ the FSU2H parameter set~\cite{Tolos:2016hhl,Tolos:2017lgv}, which reproduces nuclear and finite-nuclei properties~\cite{Chen:2014sca, Danielewicz:2002pu, Fuchs:2000kp, Lynch:2009vc} while accommodating 2$M_\odot$ neutron stars and radii constraints inferred from the GW170817~\cite{Fattoyev:2017jql, Annala:2017llu, Krastev:2018nwr, Most:2018hfd, LIGOScientific:2017vwq}. 

The thermodynamic potential and the mesonic equations of motion for both models are presented in App.~\ref{sec:app_formalism}. In App.~\ref{sec:app_hadron_params}, we describe the relations between hyperon vector couplings and nucleon couplings, and summarize the model parameters and some properties of the NL$\omega\rho$ and FSU2H models in Table~\ref{tab.Params_nlw_fus2h}.

\subsection{Mixed phase}
The chemical potential of particle $i$ in phase $a$ ($a\in\{Q,H\}$) can be written as the linear combination of the chemical potentials associated with the three conserved charges,
\begin{equation}
    \mu_i^a = B_i\mu_B + I_{3i}\mu_3 + S_i\mu_S, 
\end{equation}
where $B_i$, $I_{3i}$ and $S_i$ are the baryon number, the third component of isospin, and the strangeness of particle $i$, respectively, and $\mu_B$, $\mu_3$ and $\mu_S$ are the corresponding chemical potentials. In the quark phase, the quark chemical potentials are related to the baryon, isospin and strangeness chemical potentials defined by
\begin{equation}
\begin{gathered}
    \mu^Q_B = \frac{3}{2}(\mu_u + \mu_d), \quad \mu_3^Q = \mu_u - \mu_d, \\
    \mu^Q_S = \frac{1}{2}(\mu_u + \mu_d - 2\mu_s).
\end{gathered}
\end{equation}
Analogous definitions hold for hadronic matter:
\begin{equation}
\begin{gathered}
    \mu_B^H = \frac{1}{2}(\mu_p + \mu_n),\quad \mu_3^H = \mu_p - \mu_n, \\
    \mu_S^H = \frac{1}{2}(\mu_p + \mu_n - 2\mu_{\Lambda}).
\end{gathered}
\end{equation}
The chemical potential of other baryons in the baryon octet are written in terms of $\mu_n$, $\mu_p$ and $\mu_{\Lambda}$:
\begin{equation}
\begin{aligned}
    \mu_{\Sigma^+} &= \mu_p - \mu_n + \mu_{\Lambda}, & \mu_{\Xi^0}  &= 2\mu_{\Lambda} - \mu_n, \\
    \mu_{\Sigma^0} &= \mu_{\Lambda},                & \mu_{\Xi^-}  &= 2\mu_{\Lambda} - \mu_p, \\
    \mu_{\Sigma^-} &= \mu_n - \mu_p + \mu_{\Lambda}.
\end{aligned}
\end{equation}
 One obtains the density of a total conserved charge ${\cal Q}$ in phase $a$ from $\rho^a_{\cal Q} = \sum_i {\cal Q}_i\rho_i$, where ${\cal Q}_i$ is the corresponding quantum number, and for a vanishing net strangeness in both phases.
 
\begin{figure*}[]
	\centering
	\begin{subfigure}[b]{0.45\linewidth}
		\centering
        \includegraphics[width=\linewidth]{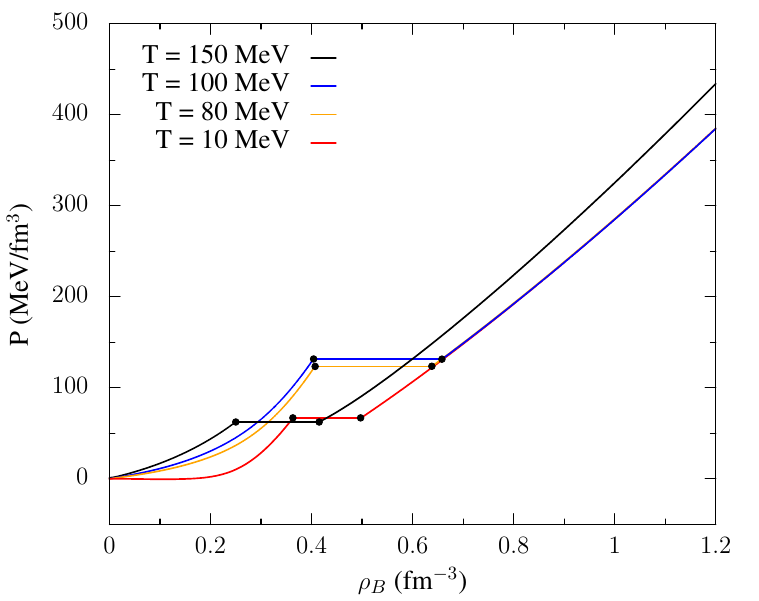}
		\caption{$\zeta = 0$ and $\alpha = 0$}
        \label{fig:isotherms_nl3wr_a}
	\end{subfigure}
	\hfil
	\begin{subfigure}[b]{0.45\linewidth}
		\centering
		\includegraphics[width=\linewidth]{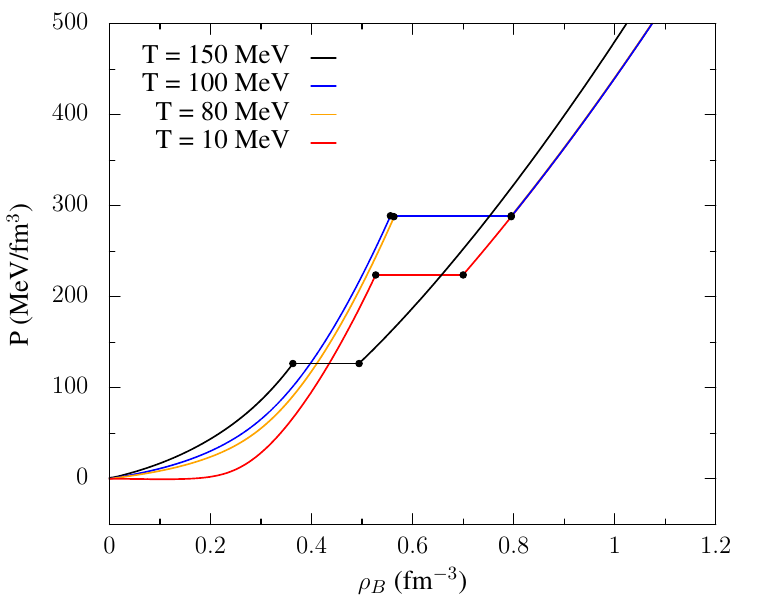}
		\caption{$\zeta = 0.5$ and $\alpha = 0$}
        \label{fig:isotherms_nl3wr_b}
	\end{subfigure}
	\begin{subfigure}[b]{0.45\linewidth}
		\centering
        \includegraphics[width=\linewidth]{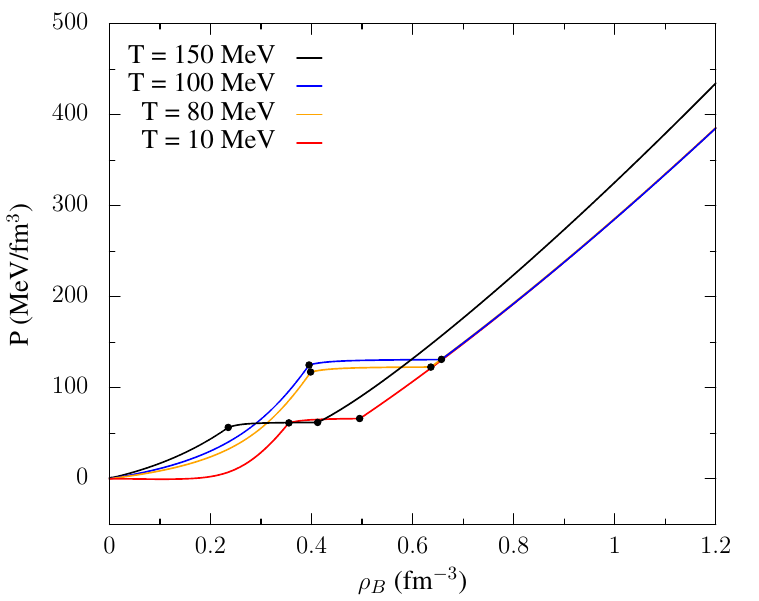}
		\caption{$\zeta = 0$ and $\alpha = 0.2$}
        \label{fig:isotherms_nl3wr_c}
	\end{subfigure}
\hfil
	\begin{subfigure}[b]{0.45\linewidth}
		\centering
		\includegraphics[width=\linewidth]{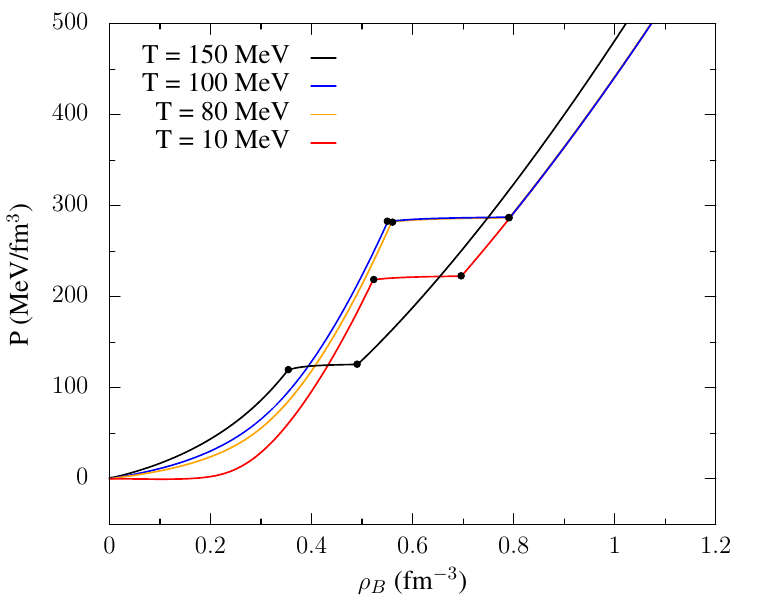}
		\caption{$\zeta = 0.5$ and $\alpha = 0.2$}
        \label{fig:isotherms_nl3wr_d}
	\end{subfigure}
	
	\caption{{Phase transition in the $P-\rho_B$ plane for symmetric ($\alpha = 0$, top) and asymmetric ($\alpha = 0.2$, bottom) matter, computed with the NL3$\omega\rho$-PNJL two-model approach. Columns show scenarios with $\zeta = 0$ (left) and $\zeta = 0.5$ (right).  Solid lines are isotherms at fixed temperatures $T =$ 10, 80, 100, 150 MeV. The region between the black dots indicates the mixed (coexistence) phase.}}
	\label{fig:isotherms_nl3wr}
\end{figure*}

\begin{table*}[t!]
    \centering
    \begin{tabular}{c c c | c c c c c|| c c c c c}
    \hline
        \multirow{2}{*}{$\zeta$} && \multirow{2}{*}{$\alpha$} &$\rho^i_B$ & $\rho^f_B$& $\Delta\rho_B$ & $P^i$ & $P^f$ &  $\rho^i_B$ & $\rho^f_B$ & $\Delta\rho_B$ & $P^i$ & $P^f$ \\
         &&&  (fm$^{-3}$) & (fm$^{-3}$)& (fm$^{-3}$) & (MeV/fm$^{3}$) & (MeV/fm$^{3}$)    & (fm$^{-3}$) & (fm$^{-3}$) & (fm$^{-3}$)  & (MeV/fm$^{3}$) & (MeV/fm$^{3}$) \\
    \hline
    \multicolumn{3}{c}{} &\multicolumn{5}{c||}{$T = 10$ MeV} & \multicolumn{5}{c}{$T = 80$ MeV} \\
    \hline
    0   && 0    & 0.363  & 0.497 & 0.134 & 67.0   & 67.0    &   0.407 &  0.638 & 0.231 & 123.5 & 123.5 \\ 
    0.5 && 0    & 0.527  & 0.700  & 0.173 & 224.0  & 224.0  &   0.563 & 0.795  & 0.232 & 288.0   & 288.0  \\
    0   && 0.2  & 0.355  & 0.495 & 0.140 & 61.5 & 66.3   &  0.398 & 0.636  & 0.238 & 117.4 & 122.8  \\
    0.5 && 0.2  & 0.523  & 0.696 & 0.173  & 219.0  & 223.0   &    0.560  & 0.790   & 0.230 & 282.0   & 287.0 \\
    \hline 
    \multicolumn{3}{c}{} &\multicolumn{5}{c||}{$T = 100$ MeV} & \multicolumn{5}{c}{$T = 150$ MeV} \\
    \hline
    0   && 0   & 0.404 &  0.658 & 0.254 & 131.5 & 131.5   &   0.250  & 0.415 & 0.165 & 62.5 & 62.5 \\
    0.5 && 0   & 0.556 & 0.795  & 0.239 & 289.0 & 289.0     & 0.363  & 0.494 & 0.131 & 126.7 & 126.7  \\
    0   && 0.2 & 0.395 & 0.657  & 0.262 & 125.1 & 131.3 &   0.235  & 0.412 & 0.177 & 56.5 & 62.0 \\
    0.5 && 0.2 & 0.550  & 0.790   & 0.240 & 283.0 &  287.0    &  0.354  & 0.490  & 0.136 & 120.0   & 126.0  \\
    \hline
    \end{tabular}
    \caption{{Initial and final boundaries of the mixed phase in the $P-\rho_B$ plane along isothermal trajectories $T=$ 10, 80, 100, 150 MeV, for the scenarios shown in Fig.~\ref{fig:isotherms_nl3wr}, within the NL3$\omega\rho$-PNJL two-phase model. The interval $\Delta\rho_B$ is defined as $\Delta\rho_B = \rho_B^f - \rho_B^i$.}}
    \label{tab:isotherms_nl3wr}
\end{table*}

In the mixed (coexistence) phase of quarks and hadrons, the Gibbs conditions (thermal, chemical and mechanical equilibrium)
\begin{subequations}\label{gibbs}
    \begin{align}
        P^H(\rho_B, \rho_3, T) &= P^Q(\rho_B, \rho_3, T), \label{gibbs_P} \\
        \mu^Q_B(\rho_B, \rho_3, T) &= \mu_B^H(\rho_B, \rho_3, T), \label{gibbs_baryon} \\
        \mu_3^Q(\rho_B, \rho_3, T) &= \mu^H_3(\rho_B, \rho_3, T),\label{gibbs_isospin}
    \end{align}
\end{subequations}
should be fulfilled.

These equilibrium conditions describe the hadron-quark mixed phase. These conditions impose global instead of local conservation conditions; therefore, the coexisting hadronic and quark phases are allowed to be charged separately.
In Eqs.~\eqref{gibbs}, the total baryon density in the mixed phase can be defined as
\begin{equation}
    \rho_B = (1 - \lambda)\rho^H_B + \lambda\rho^Q_B,
\end{equation}
where $\lambda$ is the quark fraction, and the total isospin density as 
\begin{equation}
    \rho_3 = (1 -\lambda)\rho_3^H + \lambda \rho_3^Q.
\end{equation}
Similarly, the total mixed phase energy density is given by
\begin{equation}
    \mathcal{E} = (1- \lambda) \mathcal{E}^H + \lambda \mathcal{E}^Q\,.
\end{equation}
For HICs with a fixed global isospin asymmetry, the global asymmetry parameter  $\alpha$ in the coexistence phase
\begin{equation}
    \alpha = -2\frac{\rho_3}{\rho_B} = -2\frac{(1 -\lambda)\rho_3^H + \lambda \rho_3^Q}{(1 -\lambda)\rho_B^H + \lambda \rho_B^Q},
\end{equation}
should be constant according to the isospin charge conservation in strong interaction. {The values of $\alpha$ are tabulated for several heavy-ion sources in Ref.~\cite{Cavagnoli:2010yb}; for stable nuclei the largest value is $\alpha = 0.227$ in $^{238}$U + $^{238}$U collisions. For this reason, we adopt $\alpha = 0.2$ to study the effect of isospin asymmetry on the phase transition and thermodynamic properties along isentropic lines. We remark that collisions involving neutron-rich unstable nuclei may reach still larger asymmetries.
}

In addition, we impose vanishing net strangeness in each phase, since the colliding nuclei carry no net strangeness~\cite{Hempel:2013tfa}. Accordingly, we set the strangeness fraction $Y_S = \rho_S/\rho_B$ to zero in both phases,
\begin{equation}
    Y_S^Q  = Y_S^H = 0,
\end{equation}
which requires the strange chemical potential to remain finite and to differ between the two phases, $\mu_S^Q \neq \mu_S^H$.


\section{Results and Discussion}\label{sec:Results}

Now, we present and discuss the results obtained considering several temperatures, isentropic lines and isospin asymmetries. In particular, we discuss the effect of introducing the vector interaction in the quark sector and  {identify the locations  of various quark fractions in the mixed phase.} {In Sec.~\ref{woHyperons} only nucleons are included in the hadronic phase, whereas the onset and effects of hyperons are treated in Sec.~\ref{wHyperons}.}

\subsection{Results without hyperons} \label{woHyperons}

The equation of state was calculated imposing Gibbs conditions defined in Eq.~(\ref{gibbs}) to determine the mixed phase.
Fig.~\ref{fig:isotherms_nl3wr} shows the pressure as a function of the baryon density to illustrate the hadron-quark phase transition: the panels illustrate two scenarios for the quark vector interaction,  $\zeta=0$ (left) and $\zeta = 0.5$ (right),  and two different isospin asymmetries, $\alpha = 0$ (top) and $\alpha = 0.2$ (bottom). The transition lines were obtained at fixed temperatures $T =$ 10, 80, 100, 150 MeV, chosen to probe both high baryon density effects and the region close to the CEP. Black dots denote the limits of the mixed phase region.

{In the isospin-symmetric matter ($\alpha = 0$) case, Figs.~\ref{fig:isotherms_nl3wr_a},~\ref{fig:isotherms_nl3wr_b}, pressure remains constant throughout the mixed phase, as expected for a Maxwell construction, even though more than one globally conserved charge is present. Since the isospin chemical potential $\mu_3 = 0$ (i.e. $\mu_u = \mu_d$ and $\mu_p =  \mu_n$) in symmetric matter, Eq.~\eqref{gibbs_isospin} is automatically satisfied, and only Eq.~\eqref{gibbs_baryon} is relevant, and the Maxwell and Gibbs constructions are equivalent in this case. Hence no isospin distillation occurs, i.e., there is no transfer of isospin per baryon, between the two phases~\cite{Hempel:2013tfa}. At variance in the asymmetric case ($\alpha=0.2$), Figs.~\ref{fig:isotherms_nl3wr_c},~\ref{fig:isotherms_nl3wr_d}, pressure increases monotonically with density.}

\begin{figure*}[htpb]
\centering
	\begin{subfigure}[b]{0.45\linewidth}
		\centering
		\includegraphics[width=\linewidth]{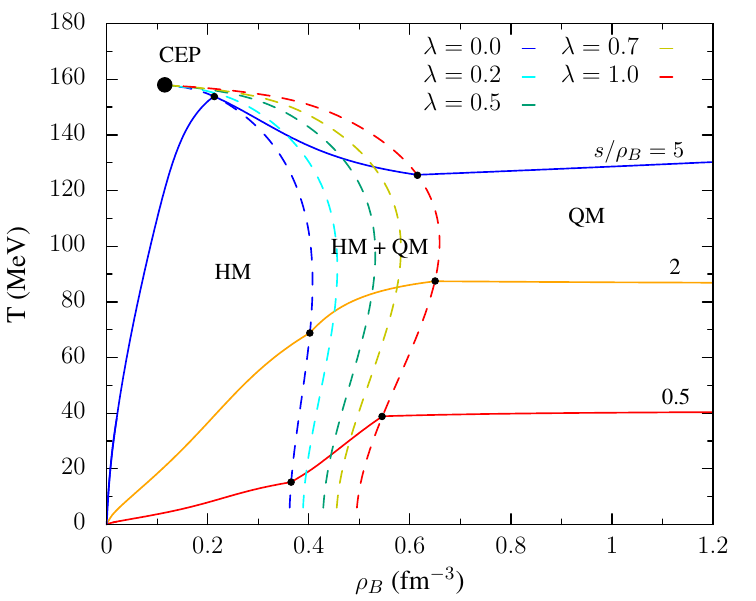}
		\caption{$\zeta = 0$ and $\alpha = 0$}
        \label{fig:T-rho_nl3wr_a}
	\end{subfigure}
	\hfil
	\begin{subfigure}[b]{0.45\linewidth}
		\centering
		\includegraphics[width=\linewidth]{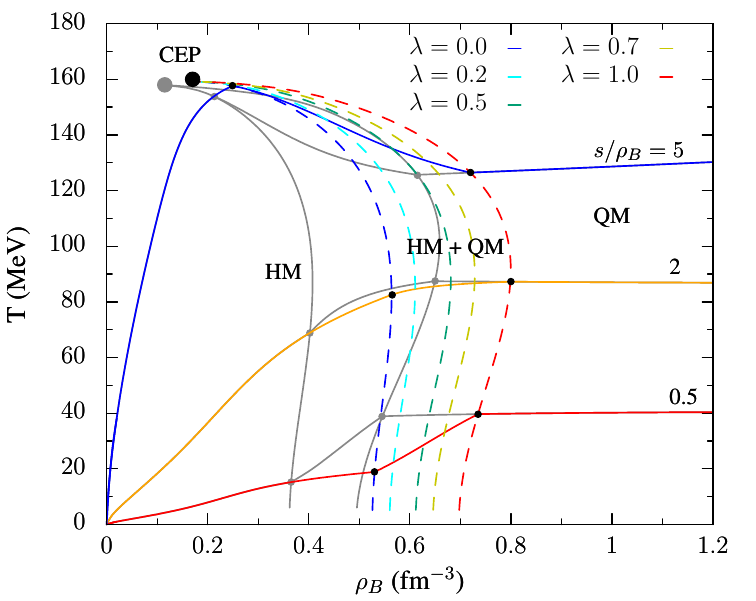}
		\caption{$\zeta = 0.5$ and $\alpha = 0$}
        \label{fig:T-rho_nl3wr_b}
	\end{subfigure}
	\begin{subfigure}[b]{0.45\linewidth}
		\centering
		\includegraphics[width=\textwidth]{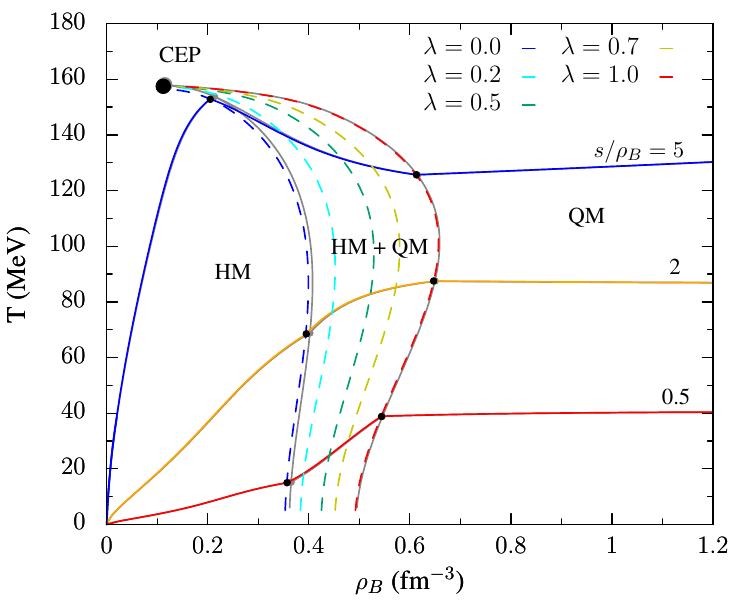}
		\caption{$\zeta = 0$ and $\alpha = 0.2$}
        \label{fig:T-rho_nl3wr_c}
	\end{subfigure}
	\hfil
	\begin{subfigure}[b]{0.45\linewidth}
		\centering
		\includegraphics[width=\linewidth]{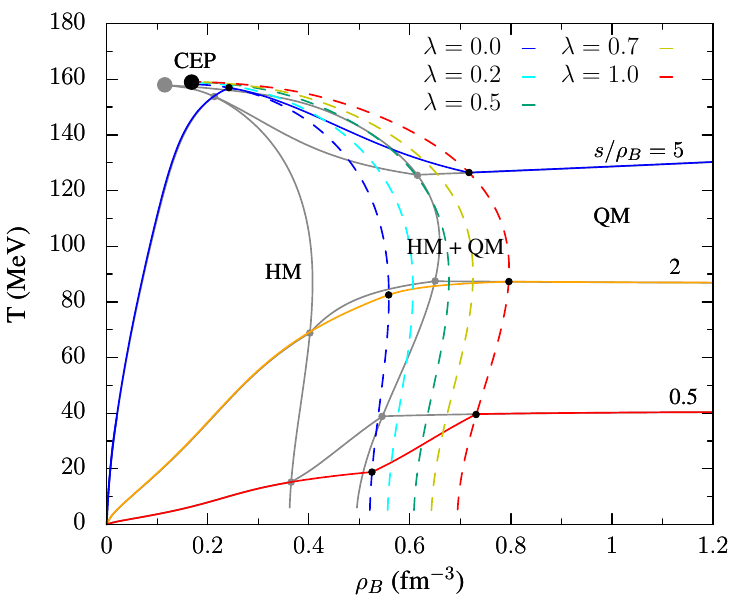}
		\caption{$\zeta = 0.5$ and $\alpha = 0.2$}
        \label{fig:T-rho_nl3wr_d}
	\end{subfigure}
	
    \caption{Phase diagram of the NL3$\omega\rho$-PNJL two-phase model in the $T-\rho_B$ plane for $\alpha = 0$ (top) and $\alpha = 0.2$ (bottom) matter, and $\zeta = 0$ (left) and $\zeta = 0.5$ (right). The big black dot is the CEP and the colored dashed lines identify the coexistence phase where $0\%$ (pure-hadron), $20\%$, $50\%$, $70\%$, and $100\%$  (pure-quark) of  matter is in the quark phase, respectively. The solid colored lines are isentropes for fixed $s/\rho_B =$ 0.5, 2, 5 ratios. For clarity, the pure-hadron and pure-quark boundaries, together with the isentropes from panel~\ref{fig:T-rho_nl3wr_a}, are displayed in gray in the other three panels to highlight differences. }
	\label{fig:T-rho_nl3wr}
\end{figure*}

The inclusion of the repulsive vector interaction in the quark sector ($\zeta \neq 0$) stiffens the equation of state: the transition pressure increases by more than a factor of two at all temperatures compared with the $\zeta = 0$ cases, and the onset of the phase transition shifts to higher baryon densities. Table~\ref{tab:isotherms_nl3wr} shows that the effect of the vector interaction on the transition strength depends on temperature. {For example, comparing the cases $\zeta = 0$ and $\zeta = 0.5$ with $\alpha = 0$, at $T = 10$ MeV, the phase transition is strengthened by including vector interactions, i.e., the density jump $\Delta\rho_B$ increases. In contrast, at $T=150$ MeV the transition is weakened, and $\Delta\rho_B$ is reduced.} This behavior reflects the nonmonotonic structure of the phase diagram and the influence of the CEP location in modifying the size of the coexistence region, as shown in Figs.~\ref{fig:T-rho_nl3wr_a} and \ref{fig:T-rho_nl3wr_b}. 

In Fig.~\ref{fig:T-rho_nl3wr} the boundaries of the hadron-quark transition in the $T-\rho_B$ plane are shown for the same set of scenarios discussed above. Dashed colored lines indicate matter with different quark fractions, $\lambda$: $\lambda = 0$ (at low density) corresponds to the pure-hadron phase boundary, while $\lambda = 1$ (at high density) corresponds to the pure-quark phase boundary; intermediate values $0<\lambda<1$ mark the mixed phase region.

At low $T$, the transition sets in slightly earlier for isospin-asymmetric matter ($\alpha = 0.2$), bottom panels, than for symmetric matter ($\alpha = 0$), top  panels; the same trend is observed in the isotherms listed in Table~\ref{tab:isotherms_nl3wr}. This behavior reflects the reduced binding of asymmetric matter, which favors the appearance of quark matter at lower baryon density. The effects of vector interactions are most pronounced at low $T$ ({compare Figs.~\ref{fig:T-rho_nl3wr_a},~\ref{fig:T-rho_nl3wr_c} with Figs.~\ref{fig:T-rho_nl3wr_b},~\ref{fig:T-rho_nl3wr_d}}): they shift the onset density upward by a factor of about $1.4$ and enlarge the mixed phase region, owing to the repulsive contribution from the isoscalar-vector channel. Similar shifts have  been reported previously (see, e.g., Refs.~\cite{Shao:2012tu, Shao:2013toa}).

\begin{table}[]
   
    \centering
    \begin{tabular}{c c c | c c c c c }
    \hline
        $\zeta$ && $\alpha$ & $T^{\rm CEP}$ (MeV) &&  $\rho_B^{\rm CEP}$ (fm$^{-3})$ && $\mu_B^{\rm CEP}$ (MeV) \\
    \hline
    0   && 0    & 158.0   && 0.115 && 515.0 \\
    0.5 && 0    & 160.0   && 0.170 && 606.7 \\
    0   && 0.2  & 157.5   && 0.112 && 508.0 \\
    0.5 && 0.2  & 159.0   && 0.168 && 604.2 \\
    \hline
    \end{tabular}
 \caption{Temperature, baryon density, and baryon chemical potential for the CEP in the different scenarios considered within the NL3$\omega\rho$-PNJL model.}
    \label{tab:cep_nl3wr}
\end{table}

{The CEP is present in all scenarios considered (Fig.~\ref{fig:T-rho_nl3wr}). For symmetric matter, Figs.~\ref{fig:T-rho_nl3wr_a},~\ref{fig:T-rho_nl3wr_b}, the coexistence boundaries corresponding to $\lambda = 0$ and $\lambda = 1$ share a common endpoint at high temperature. In asymmetric matter Figs.~\ref{fig:T-rho_nl3wr_c},~\ref{fig:T-rho_nl3wr_d}, the endpoints of the $\lambda = 0$ and $\lambda = 1$ lines are slightly separated, so that each fixed value of $\lambda$ admits its own endpoint inside the mixed region. As explained in Ref.~\cite{Shao:2016fsh}, for a given $\lambda$, one has $\rho_B^H  < \rho_B^Q$ in the coexistence region and the two curves $T(\rho_B^H)$ and $T(\rho_B^Q)$ intersect at single point at high $T$ -- the CEP for that $\lambda$. Thus, asymmetric matter exhibits a $\lambda$-dependent set of critical points; linking these endpoints produces a short transition line between the $\lambda = 0$ and $\lambda =1$ boundaries. For $\alpha = 0.2$ the temperature difference between these endpoints is small ($\approx1$ MeV), so we adopt the midpoint point of the line connecting both the extreme of the $\lambda = 0$ and $\lambda =1$ boundaries at high $T$ as the representative CEP in asymmetric scenarios.} Table~\ref{tab:cep_nl3wr} summarizes the CEP positions predicted by the NL3$\omega\rho$-PNJL scheme for each scenario. The CEP temperature $T^{\rm CEP}$ is largely insensitive to the cases considered, although it is systematically lower for isospin-asymmetric matter than for the symmetric case. However, the CEP baryon density  $\rho_B^{\rm CEP}$ and baryon chemical potential $\mu_B^{\rm CEP}$ exhibit more visible changes. Isospin asymmetry tends to shift the CEP to lower densities and chemical potentials, but the inclusion of repulsive vector interactions has the dominant (opposite) effect. 
  
Figure~\ref{fig:T-rho_nl3wr} also shows isentropic trajectories at fixed $s/\rho_B =$ 0.5, 2, 5, where the last value was chosen to approach the CEP as closely as possible before numerical instabilities associated with the critical point appear. As expected, in all scenarios the isentropes approach the vacuum limit as $T\to 0$ (i.e., $\rho_B \to 0$ and $\mu_B \to m_N$). This behavior follows from the third law of thermodynamics combined with the constraint of fixed $s/\rho_B$. 

\begin{table}[htpb]
    \centering
    \begin{tabular}{c c c | c c c c c c}
    \hline
        \multirow{2}{*}{$\zeta$} && \multirow{2}{*}{$\alpha$} &$\rho^i_B$ & $\rho^f_B$ & $T^i$ & $T^f$ & $\Delta\rho_B$ & $\Delta T$ \\
         &&&  (fm$^{-3}$) & (fm$^{-3}$)& (MeV) & (MeV) & (fm$^{-3}$)    & (MeV) \\
    \hline
    \multicolumn{3}{c}{} &\multicolumn{6}{c}{$s/\rho_B = 0.5$} \\
    \hline
    0   && 0    & 0.365 & 0.545 & 15.2 & 38.7 & 0.180 & 23.5  \\ 
    0.5 && 0    & 0.530 & 0.735 & 18.9 & 39.6 & 0.205 & 20.7  \\
    0   && 0.2  & 0.357 & 0.544 & 15.0 & 38.8 & 0.187 & 23.8   \\
    0.5 && 0.2  & 0.525 & 0.731 & 18.8 & 39.5 & 0.206 & 20.7  \\
    \hline 
    \multicolumn{3}{c}{} &\multicolumn{6}{c}{$s/\rho_B = 2$} \\
    \hline
    0   && 0    & 0.402 & 0.650 & 68.8 & 87.5 & 0.248 & 18.7   \\ 
    0.5 && 0    & 0.565 & 0.800 & 82.5 & 87.3 & 0.235 & 4.8  \\
    0   && 0.2  & 0.395 & 0.647 & 68.5 & 87.5 & 0.252 & 19.0  \\
    0.5 && 0.2  & 0.558 & 0.796 & 82.5 & 87.3 & 0.237 & 4.8   \\
    \hline
    \multicolumn{3}{c}{} &\multicolumn{6}{c}{$s/\rho_B = 5$} \\
    \hline
    0   && 0    & 0.213  & 0.615  & 153.8  & 125.5  & 0.402 & -28.3   \\ 
    0.5 && 0    & 0.244  & 0.720  & 157.7  & 126.5  & 0.476 & -31.2  \\
    0   && 0.2  & 0.205  & 0.613  & 152.8  & 125.7  & 0.408 & -27.1   \\
    0.5 && 0.2  & 0.242  & 0.717  & 157.0  & 126.5  & 0.475 & -30.5   \\
    \hline
    \end{tabular}
    \caption{Limits of the mixed phase in the $T-\rho_B$ plane along the isentropes $s/\rho_B = 0.5$, 2, 5 for the different scenarios shown in Fig.~\ref{fig:T-rho_nl3wr}. Here, the intervals $\Delta\rho_B$ and $\Delta T$ are defined as $\Delta\rho_B = \rho_B^f - \rho_B^i$ and $\Delta T = T^f - T^i$, respectively.}
    \label{tab:isentropes_nl3wr}
\end{table}

As $\rho_B$ increases and the isentropes cross the pure-hadron boundary ($\lambda = 0$), their evolution inside the mixed phase depends on the value of $s/\rho_B$, producing a nonmonotonic behavior. In Fig.~\ref{fig:T-rho_nl3wr_a}, the temperature  for the $s/\rho_B = 5$ trajectory  drops as the quark fraction increases inside the mixed; the trajectory then leaves the mixed phase at $\lambda=1$, after which $T$ slowly increases. By contrast, along the $s/\rho_B = 0.5$ and $s/\rho_B = 2$ trajectories the system temperature rises while in the mixed phase. 
An identical trend is observed when $\zeta = 0.5$ and/or $\alpha = 0.2$. Table~\ref{tab:isentropes_nl3wr} and Fig.~\ref{fig:T-rho_nl3wr_b} show that including repulsive vector interactions increases the values of $T$ and $\rho_B$ at the onset of the mixed phase, i.e., at the $\lambda=0$ boundary (cf. Table~\ref{tab:isentropes_nl3wr}). Also, each isentrope leaves the mixed phase at a significantly higher density relative to the scenario in Fig.~\ref{fig:T-rho_nl3wr_a}, but with approximately the same temperature. By comparison, the effect of an asymmetry parameter of  $\alpha = 0.2$ (Fig.~\ref{fig:T-rho_nl3wr_c}) is much smaller, as the temperature and density at the intersection with the $\lambda = 0$ boundary remain mostly unchanged. The asymmetry $\alpha = 0.2$ is still too small to produce significant changes in the phase boundaries and isentropes. In particular, the pure-quark boundary nearly overlaps with the symmetric case (Fig.~\ref{fig:T-rho_nl3wr_a}). We note that, at the mixed phase quark phase transition, chiral symmetry is already largely restored, so the light $u$ and $d$ quark masses become similar, and the isospin dependence of the EoS is small.

\begin{figure}[h]
	\centering
	\includegraphics[width=\linewidth]{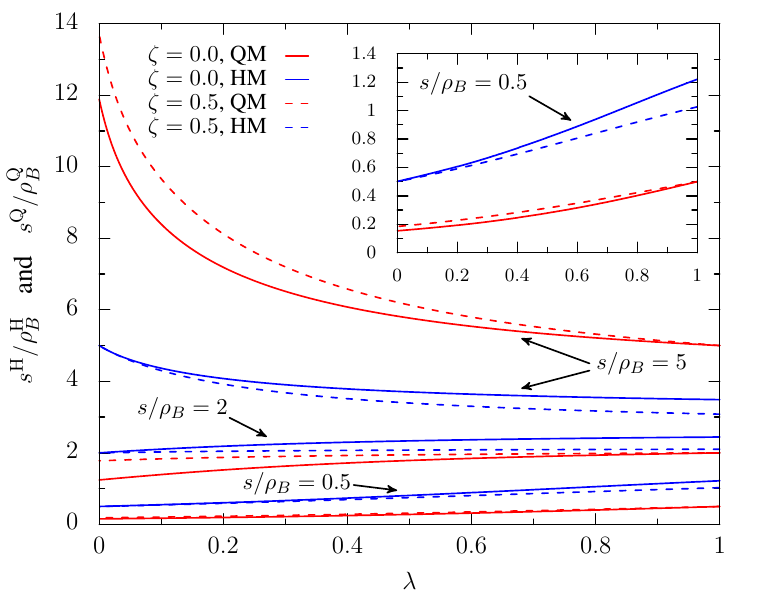}
	\caption{Local entropy per baryon number density of quark (red) and hadron (blue) matter as a function of the quark concentration for $\zeta = 0$ (solid) and $\zeta = 0.5$ (dashed) in symmetric matter.} 
	\label{fig:s-lambda_nl3wr}
\end{figure}

The different behavior of each isentropic trajectory inside the mixed phase is better understood by looking at the $s/\rho_B$ ratio in each phase as a function of $\lambda$, as we plot in Fig.~\ref{fig:s-lambda_nl3wr}. Here, the scenarios with  isospin-symmetric matter are shown (the asymmetric case is qualitatively identical). For each isentrope, we plot $s^{\rm Q}/\rho^{\rm Q}_B$ (quark phase, red) and $s^{\rm H}/\rho^{\rm H}_B$ (hadron phase, blue).

{
    The key point is that the local entropy per baryon in each phase is generally different. To maintain a fixed global ratio $s/\rho_B$ as the quark fraction $\lambda$ changes, the system must adjust the temperature $T$ and the baryon chemical potential $\mu_B$. For the trajectories with $s/\rho_B=0.5$ and $2$ one finds $s^Q/\rho_B^Q < s^H/\rho_B^H$. This inequality reflects the dominance of density-dependent contributions over thermal ones. Since the hadron and quark Fermi energies are large compared to $T$, the entropy per baryon for a multicomponent system in a pure phase can be analyzed with the relation for degenerate Fermi particle~\cite{Steiner:2000bi}
    \begin{equation}
        \frac{s}{\rho_B} = \pi^2T \frac{\sum_i p_{F_i} \sqrt{p_{F_i}^2 + m_i^{*2}}}{\sum_i p_{F_i}^3},
    \end{equation}
    where $m^{*}_i$ and $p_{F_i}$ are the effective mass and the Fermi momentum of component $i$, respectively.
    When the isentropes cross the $\lambda=0$ boundary and enter the mixed phase, the opening of additional quarks degrees of freedom eventually lowers the baryons Fermi energy, but that reduction does not offset the larger nucleon rest mass (relative to $T$) and the comparatively small kinetic energy of the quarks. As a result, converting hadronic matter into quark matter (increasing $\lambda$) requires an increase in $T$ in order to keep the global $s/\rho_B$ fixed, which produces the observed temperature rise inside the mixed phase. By contrast, for $s/\rho_B=5$ one finds $s^Q/\rho_B^Q > s^H/\rho_B^H$. In this regime the light quark masses\footnote{Only $u$ and $d$ flavors appear in the mixed phase. In the NL3$\omega\rho$-PNJL model the $s$-quark crossover occurs only at a baryon density $\rho_B \approx 1.4$ fm$^{-3}$, which lies outside the density range displayed, and therefore its effect on the isentropes is not visible.}
    $m_{u,d}$ become comparable to $T$ (cf. Table~\ref{tab:isentropes_quark_mass_nl3wr}), and the introduction of quark degrees of freedom substantially increases the system’s specific heat because there are more particle species and the quarks behave more relativistically than the hadrons. The combination of these effects raises the entropy per baryon in the quark phase. Consequently, as the isentrope crosses the mixed phase, $T$ decreases with increasing $\lambda$, producing the observed cooling within the mixed phase. 
}
Dashed lines show the effect of vector interactions in the quark sector. Vector repulsion increases (decreases) $s^{\rm Q}/\rho^{\rm Q}_B$ ($s^{\rm H}/\rho^{\rm H}_B$) and thereby modifies the difference between the quark and hadron phase. For the $s/\rho_B = 5$ isentrope this separation grows, which explains the larger temperature variation inside the mixed phase; for  $s/\rho_B = 0.5$ and $s/\rho_B = 2$ the separation is reduced, consistent with the smaller temperature variation observed for those trajectories. In particular, the near overlap of $s^{\rm Q}/\rho^{\rm Q}_B$ and $s^{\rm H}/\rho^{\rm H}_B$ for $s/\rho_B = 2$ with $\zeta = 0.5$ accounts for the small temperature change along that isentrope.

\begin{table}[htbp]
    \centering
    \begin{tabular}{c c c | c c || c c || c c}
    \hline
        \multirow{2}{*}{$\zeta$} && \multirow{2}{*}{$\alpha$} &$m^*_u$ & $m^*_d$ &$m^*_u$ & $m^*_d$ &$m^*_u$ & $m^*_d$  \\
         &&&  (MeV) & (MeV) &  (MeV) & (MeV) &  (MeV) & (MeV) \\
    \hline
    \multicolumn{3}{c}{} &\multicolumn{2}{c||}{$s/\rho_B = 0.5$} &\multicolumn{2}{c||}{$s/\rho_B = 2$} &\multicolumn{2}{c}{$s/\rho_B = 5$} \\
    \hline
    0   && 0    & 27.8 & 27.8 & 21.6 & 21.6 & 123.4 & 123.4 \\ 
    0.5 && 0    & 17.2 & 17.2 & 15.9 & 15.9 & 140.7 & 140.7 \\
    0   && 0.2  & 33.4 & 27.8 & 26.0 & 21.2 & 152.4 & 128.5 \\
    0.5 && 0.2  & 19.0 & 16.7 & 17.7 & 15.3 & 161.5 & 142.7   \\
    \hline
    \end{tabular}
    \caption{The $u$, $d$ effective quark masses at the onset of the mixed phase along the isentropes shown in Fig.~\ref{fig:T-rho_nl3wr}.}
    \label{tab:isentropes_quark_mass_nl3wr}
\end{table}

\begin{figure*}[]
\centering
	\begin{subfigure}[b]{0.45\linewidth}
		\centering
		\includegraphics[width=\linewidth]{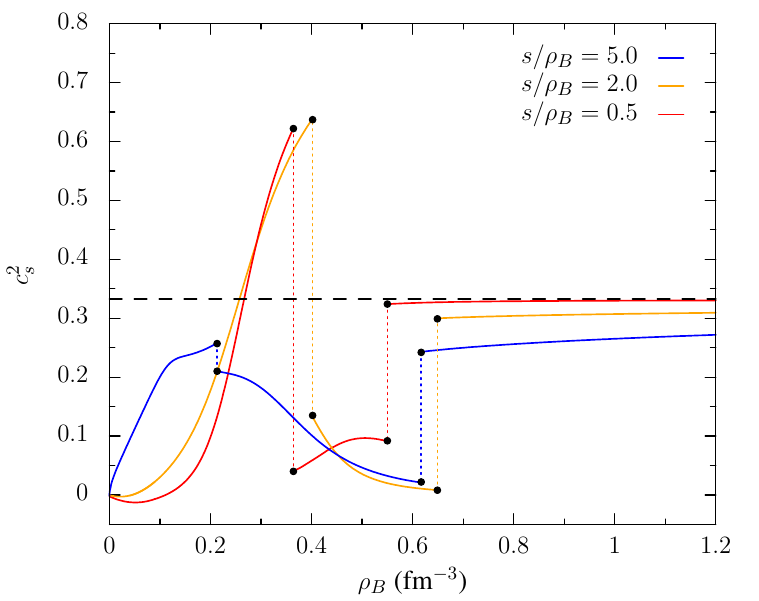}
		\caption{$\zeta = 0$ and $\alpha = 0$}
        \label{fig:vs-rho_nl3wr_a}
	\end{subfigure}
	\hfil
	\begin{subfigure}[b]{0.45\linewidth}
		\centering
		\includegraphics[width=\linewidth]{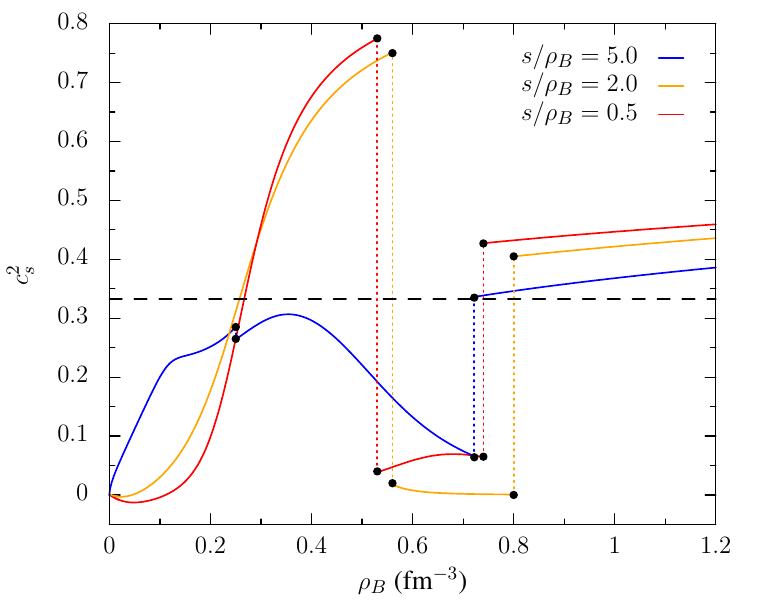}
		\caption{$\zeta = 0.5$ and $\alpha = 0$}
        \label{fig:vs-rho_nl3wr_b}
	\end{subfigure}
	\begin{subfigure}[b]{0.45\linewidth}
		\centering
		\includegraphics[width=\textwidth]{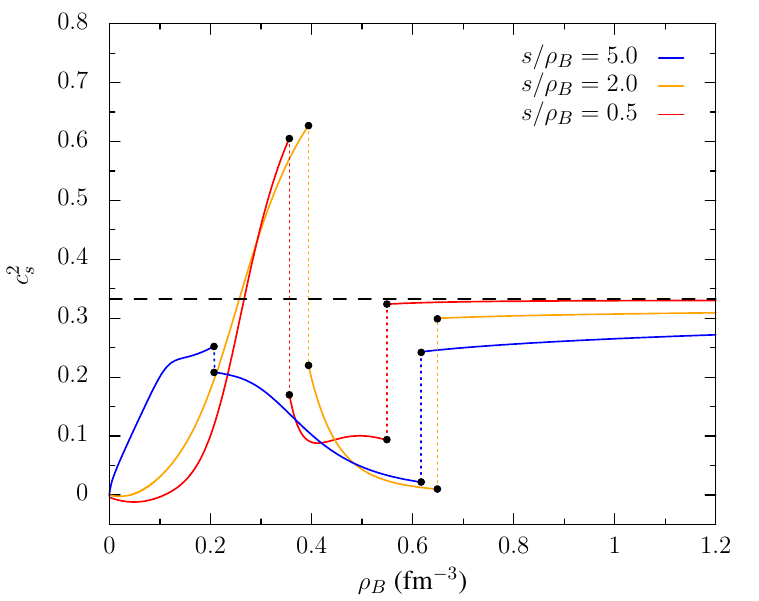}
		\caption{$\zeta = 0$ and $\alpha = 0.2$}
        \label{fig:vs-rho_nl3wr_c}
	\end{subfigure}
	\hfil
	\begin{subfigure}[b]{0.45\linewidth}
		\centering
		\includegraphics[width=\linewidth]{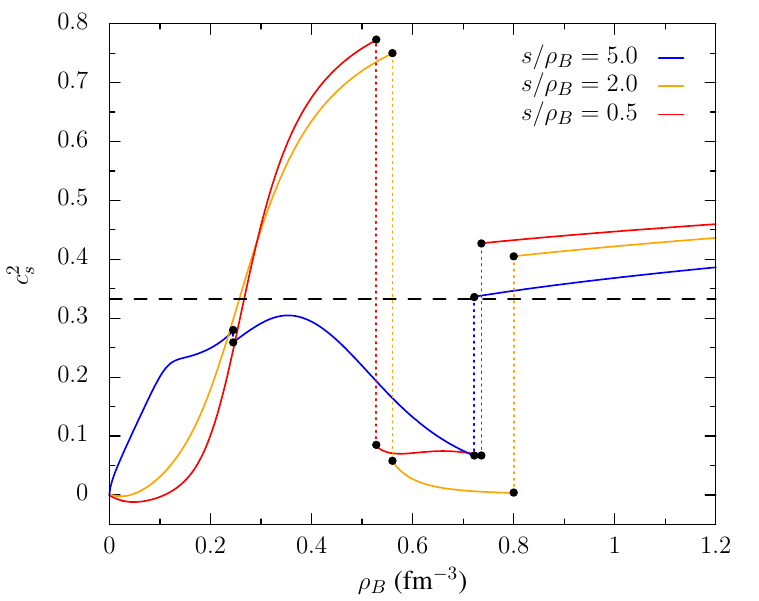}
		\caption{$\zeta = 0.5$ and $\alpha = 0.2$}
        \label{fig:vs-rho_nl3wr_d}
	\end{subfigure}
    
    \caption{Sound velocity squared $c_s ^2$ as a function of the baryon density along the isentropes $s/\rho_B =$ 0.5, 2, 5 for $\zeta = 0$ (left) and $\zeta = 0.5$ (right), and for $\alpha = 0$ (top) and $\alpha = 0.2$ (bottom) within the NL3$\omega\rho$-PNJL two-model approach. The horizontal dashed line $c_s^2 = 1/3$ indicates the high-density conformal limit.}
	\label{fig:vs-rho_nl3wr}
\end{figure*}

\begin{figure*}[t!]
\centering
	\begin{subfigure}[b]{0.45\linewidth}
		\centering
		\includegraphics[width=\linewidth]{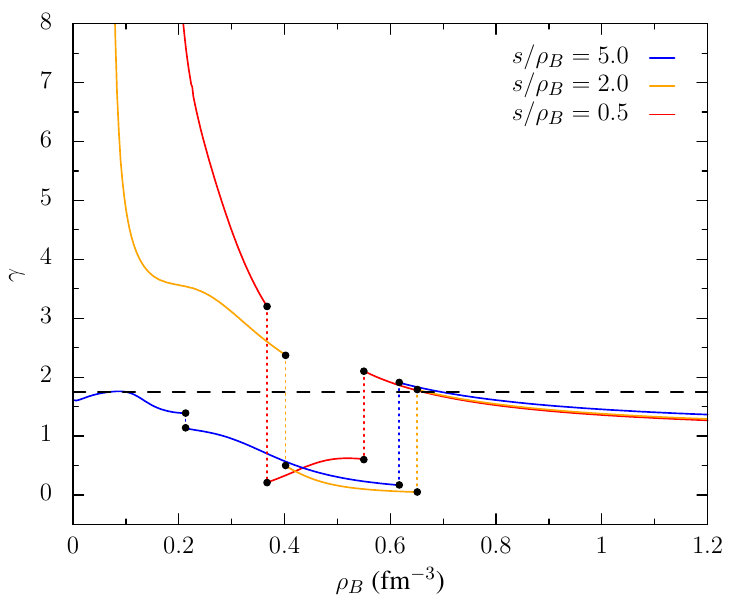}
		\caption{$\zeta = 0$ and $\alpha = 0$}
        \label{fig:gamma-rho_nl3wr_a}
	\end{subfigure}
	\hfil
	\begin{subfigure}[b]{0.45\linewidth}
		\centering
		\includegraphics[width=\linewidth]{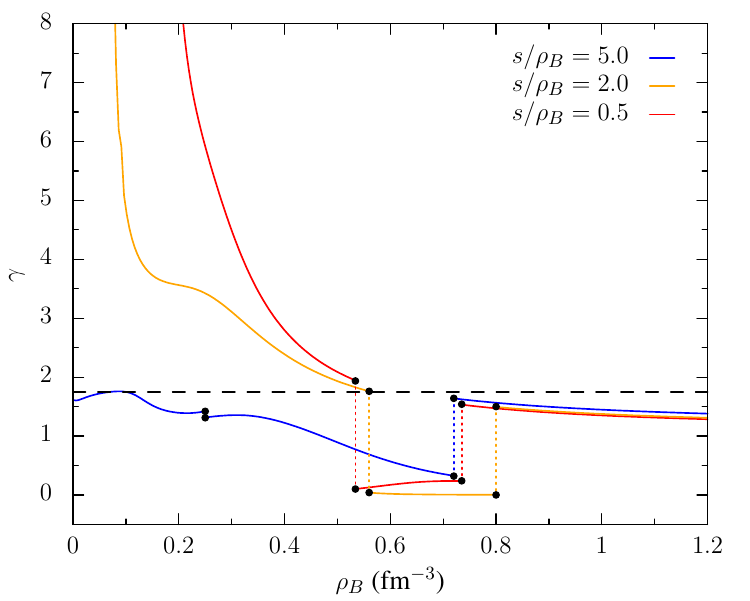}
		\caption{$\zeta = 0.5$ and $\alpha = 0$}
        \label{fig:gamma-rho_nl3wr_b}
	\end{subfigure}
	\begin{subfigure}[b]{0.45\linewidth}
		\centering
		\includegraphics[width=\textwidth]{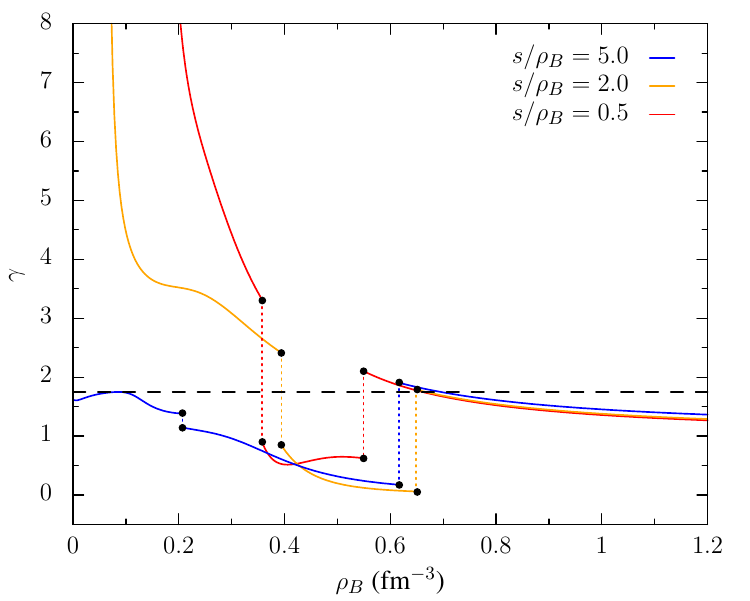}
		\caption{$\zeta = 0$ and $\alpha = 0.2$}
        \label{fig:gamma-rho_nl3wr_c}
	\end{subfigure}
	\hfil
	\begin{subfigure}[b]{0.45\linewidth}
		\centering
		\includegraphics[width=\linewidth]{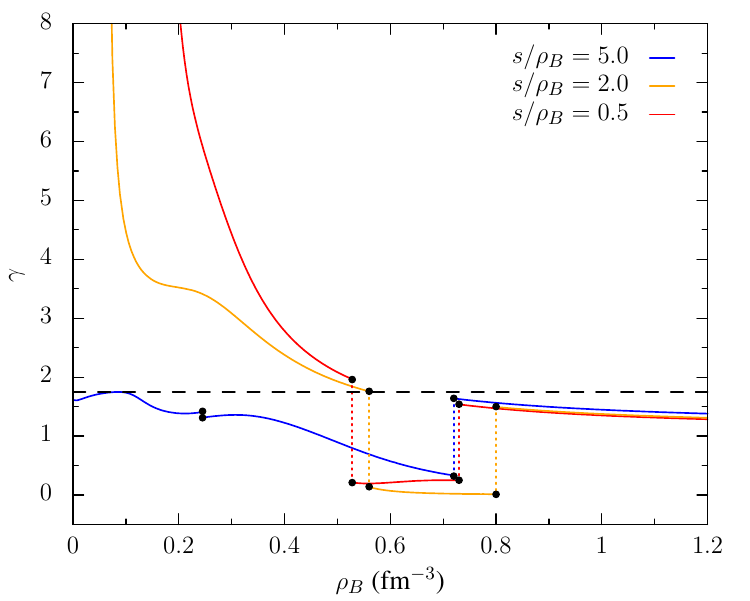}
		\caption{$\zeta = 0.5$ and $\alpha = 0.2$}
        \label{fig:gamma-rho_nl3wr_d}
	\end{subfigure}
	
 \caption{Polytropic index $\gamma$ as a function of the baryon density along the isentropes $s/\rho_B =$ 0.5, 2, 5 for $\zeta = 0$ (left) and $\zeta = 0.5$ (right), and for $\alpha = 0$ (top) and $\alpha = 0.2$ (bottom) within the NL3$\omega\rho$-PNJL two-model approach. The horizontal dotted line $\gamma = 1.75$ serves as an approximate reference value to distinguish hadronic from quark degrees of freedom.}
 \label{fig:gamma-rho_nl3wr}
\end{figure*}

The nonmonotonic evolution of isentropes across the phase diagram can leave measurable signatures on thermodynamic observables in HICs. The sound velocity, $c_s^2 = \left({\rm d} P/{\rm d}  \mathcal{E}\right)_{s/\rho_B}$, controls the hydrodynamic response to initial energy-density gradients and thus influences the collective acceleration of the fireball~\cite{Motta:2020cbr}. Because $c_s^2$ changes during the expansion, notably on entering or leaving the mixed phase, these changes may leave an imprint on the transverse-momentum distributions of hadrons at freeze-out~\cite{Mohanty:2003va}. Experimentally, $c_s^2$ can be extracted from the widths of rapidity (or pseudo-rapidity) distributions~\cite{Mohanty:2003va, Gao:2015sdb, ALICE:2016fbt}. Another useful  parameter is the polytropic index~\cite{Annala:2019puf} $\gamma = \left({\rm d}\ln P/{\rm d}\ln\cal{E}\right)_{s/\rho_B}$, recently employed with the PNJL model~\cite{Liu:2023ocg} and within the context of quark-matter cores in hybrid stars in Ref.~\cite{Liu:2023gmq}. In the conformal limit $\gamma = 1$, while in chiral effective theory calculations and hadronic models generically predict $\gamma \approx 2.5$ around and above saturation density at zero temperature~\cite{Annala:2019puf, Kurkela:2009gj}.

\begin{figure*}[htpb]
	\centering
	\begin{subfigure}[]{0.45\linewidth}
		\centering
        \includegraphics[width=\linewidth]{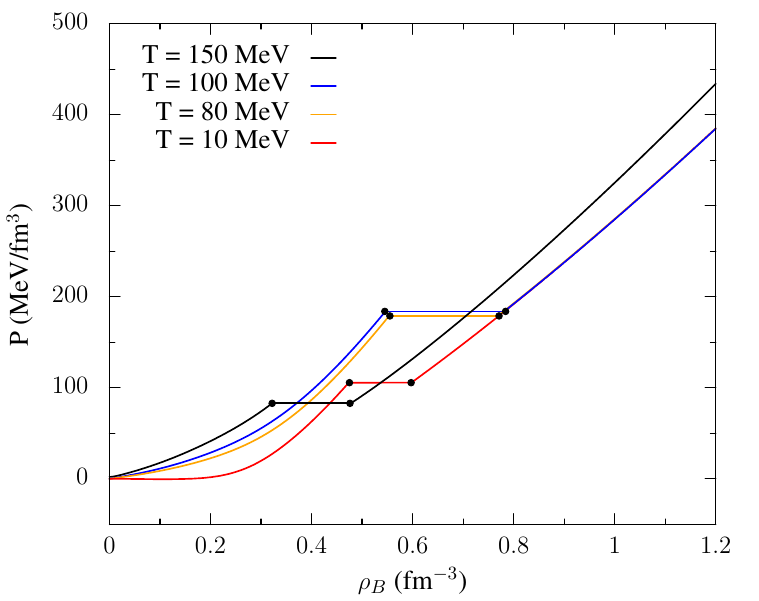}
		\caption{$\zeta = 0$ and $\alpha = 0$}
        \label{fig:isotherms_fsu2h_a}
	\end{subfigure}
    \hfil
	\begin{subfigure}[]{0.45\linewidth}
		\centering
        \includegraphics[width=\linewidth]{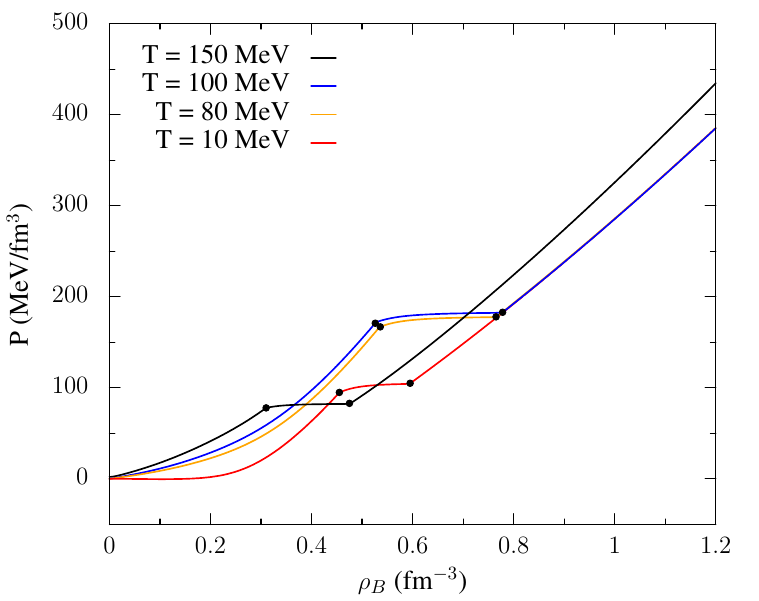}
		\caption{$\zeta = 0$ and $\alpha = 0.2$}
        \label{fig:isotherms_fsu2h_c}
	\end{subfigure}
	
	\caption{{Phase transition in the $P-\rho_B$ plane for symmetric ($\alpha = 0$, left) and asymmetric ($\alpha = 0.2$, right) matter with $\zeta = 0$, computed with the FSU2H-PNJL two-model approach. Solid lines are isotherms at fixed temperatures $T=$ 10, 80, 100, 150 MeV. The region between the black dots indicates the mixed (coexistence) phase.}}
	\label{fig:isotherms_fsu2h}
\end{figure*}

Results for $c_s^2$ along the three isentropes  are shown in Fig.~\ref{fig:vs-rho_nl3wr} as a function of $\rho_B$, for the four scenarios of previous figures. The horizontal dashed line at $c_s ^2 = 1/3$ marks the conformal limit for free massless fermions. In our results, both the $s/\rho_B = 0.5$ and $s/\rho_B = 2$ trajectories exceed this limit near $\rho_B \approx 0.21$ fm$^{-3}$, consistent with other hadronic approaches: {for instance, chiral effective field theory predicts $c_s^2 \ll 1/3$ below nuclear saturation density, while most hadronic models predict $c_s^2 \gtrsim0.5$ at higher densities~\cite{Annala:2023cwx}.} A significant feature across all scenarios is a steep drop in $c^2_s$ when an isentrope traverses the mixed phase. This behavior reflects the change in degrees of freedom: the onset of quark matter softens the equation of state and reduces $c_s^2$, which later recovers as the system reaches the pure high-density quark phase. The temporary softening reduces the fluid acceleration during the transition and may therefore leave detectable signatures in the final transverse-momentum spectra of emitted hadrons. {Additionally, the hadronic segments of the $s/\rho_B = 0.5$ and $s/\rho_B = 2$ trajectories include density intervals where $c_s^2 < 0$, corresponding to the liquid-gas phase transition of nuclear matter. For symmetric matter (Figs.~\ref{fig:vs-rho_nl3wr_a},~\ref{fig:vs-rho_nl3wr_b}) these intervals extend up to 0.111 fm$^{-3}$ and 0.045 fm$^{-3}$ for $s/\rho_B = 0.5$ and 2, respectively; for $\alpha = 0.2$ it is reduced to 0.109 fm$^{-3}$ and 0.041 fm$^{-3}$ (Figs.~\ref{fig:vs-rho_nl3wr_c},~\ref{fig:vs-rho_nl3wr_d}). Physically, $c_s^2 < 0$ implies an imaginary sound speed in the linearized hydrodynamic equation, so that small perturbations do not propagate as waves in the (unstable) spinodal region but instead exhibit exponential decay or growth (spinodal decomposition). For the $s/\rho_B = 5$ isentrope and trajectories at higher baryon density $c_s^2$ is always positive.
}

The evolution of the sound velocity inside the mixed phase depends strongly on the isentrope, $s/\rho_B$. In Fig.~\ref{fig:vs-rho_nl3wr_a}, both the $s/\rho_B = 5$ and $s/\rho_B = 2$ curves show a monotonic decrease of $c_s^2$ with increasing $\rho_B$; the steep drop at low $\rho_B$ is, however, noticeably smaller for the $s/\rho_B = 5$ case than for $s/\rho_B = 2$. By contrast, the  $s/\rho_B = 0.5$ curve, $c_s^2$ increases immediately after the discontinuity at low $\rho_B$, reaches a local maximum inside the mixed phase, and then decreases. Indeed, the same local-peak structure appears in some of the other scenarios.

Vector interactions modify these behaviors in an isentrope-dependent way as shown in Figs.~\ref{fig:vs-rho_nl3wr_b} and \ref{fig:vs-rho_nl3wr_d}. For $s/\rho_B = 5$, the discontinuity at low $\rho_B$ is reduced while increasing at high $\rho_B$ for $\zeta = 0.5$, and a local peak develops in the mixed phase. For $s/\rho_B = 0.5$ and 2, vector repulsion enhances the sound-velocity drop at both low and high densities and reduces $c_s^2$ throughout the mixed phase. We also observe that $\zeta = 0.5$ causes $c_s^2$ to exceed the conformal limit already at the mixed-quark boundary at high $\rho_B$, suggesting that such large vector couplings may be disfavored; accordingly, in the hyperon study in Sec.~\ref{wHyperons}, that follows, we adopt $\zeta = 0.1$.

Finite isospin asymmetry (Figs.~\ref{fig:vs-rho_nl3wr_c} and \ref{fig:vs-rho_nl3wr_d}) generally softens the steep changes at the mixed-phase boundaries. Its effect is largest for the $s/\rho_B = 0.5$ and $s/\rho_B = 2$  trajectories, especially at low $\rho_B$. In particular, for the curve with $s/\rho_B = 0.5$ a pronounced valley and a new local minimum appear, which are partly smoothed by quark vector interactions (compare Figs.~\ref{fig:vs-rho_nl3wr_c} and\ref{fig:vs-rho_nl3wr_d}).

\begin{table*}[]
    \centering
    \begin{tabular}{c c c | c c c c c || c c c c c }
    \hline
        \multirow{2}{*}{$\zeta$} && \multirow{2}{*}{$\alpha$} &$\rho^i_B$ & $\rho^f_B$ & $\Delta\rho_B$ & $P^i$ & $P^f$ &$\rho^i_B$ & $\rho^f_B$ & $\Delta\rho_B$ & $P^i$ & $P^f$ \\
         &&&  (fm$^{-3}$) & (fm$^{-3}$) & (fm$^{-3}$) & (MeV/fm$^{3}$) & (MeV/fm$^{3}$) & (fm$^{-3}$) & (fm$^{-3}$) &(fm$^{-3}$) & (MeV/fm$^{3}$) & (MeV/fm$^{3}$)\\
    \hline
    \multicolumn{3}{c}{} &\multicolumn{5}{c||}{$T = 10$ MeV} & \multicolumn{5}{c}{$T = 80$ MeV} \\
    \hline
    0   && 0    & 0.475  & 0.597  &0.122& 105.7   & 105.7 & 0.555 & 0.771  &0.216 & 179.0 & 179.0 \\ 
    0.1 && 0    & 0.550  & 0.680  &0.130& 155.0   & 155.0 & 0.631 & 0.844  &0.213 & 234.0 & 234.0 \\
    0   && 0.2  & 0.455  & 0.540  &0.085& 95.0    & 105.0 & 0.536 & 0.765  &0.229 & 167.0 & 178.0\\
    0.1 && 0.2  & 0.528  & 0.675  &0.147& 141.0   & 153.0 & 0.610 & 0.837  &0.227 & 219.0 & 232.0\\
    \hline 
    \multicolumn{3}{c}{} &\multicolumn{5}{c||}{$T = 100$ MeV} & \multicolumn{5}{c}{$T = 150$ MeV} \\
    \hline
    0   && 0   & 0.545 & 0.784  &0.239& 184.0 & 184.0   & 0.322  & 0.476 &0.154& 83.0 & 83.0\\
    0.1 && 0   & 0.618 & 0.845  &0.227& 235.0 & 235.0   & 0.358  & 0.500 &0.142& 99.0 & 99.0\\
    0   && 0.2 & 0.526 & 0.778  &0.252& 171.0 & 183.0   & 0.310  & 0.475 &0.165& 78.0 & 83.0\\
    0.1 && 0.2 & 0.600  & 0.843 &0.243& 221.0 & 233.0   & 0.345  & 0.496 &0.151& 93.0 & 98.0\\
    \hline
    \end{tabular}
    \caption{Initial and final boundaries of the mixed phase in the $P-\rho_B$ plane along isothermal trajectories $T=$ 10, 80, 100, 150 MeV, for the different scenarios discussed in the text, within the FSU2H-PNJL two-phase model. The interval $\Delta\rho_B$ is defined as $\Delta\rho_B = \rho_B^f - \rho_B^i$.}
    \label{tab:isotherms_fsu2h}
\end{table*}

The overall behavior of the speed of sound inside the transition region is a result of several effects, namely the entropy per baryon and energy density in each phase, the effect of the $G_V$ interaction  and the isospin asymmetry of each phase.
For $s/\rho_B=2$, both phases have similar entropy per baryon (see Fig.~\ref{fig:s-lambda_nl3wr}), and the behavior is defined by the EoS: near to the hadron transition the contribution of the stiff hadron EoS is larger, and its influence decreases as the pure-quark phase is reached, so that $c_s^2$ decreases monotonically. For $s/\rho_B=5$, close to the pure-hadron boundary, the quark phase has a much larger entropy per baryon (Fig.~\ref{fig:s-lambda_nl3wr}), so that in the mixed phase the drop of the speed of sound is not very large, and can even suffer an increase if $G_V\ne 0$, since this interaction stiffens the quark EoS. As the system moves from the pure-hadron to the pure-quark boundary, the entropy per baryon of the quark phase decreases, and the same occurs to the speed of sound. For the lowest isentrope, $s/\rho_B=0.5$, the entropy contribution of the quark phase increases as the quark fraction increases (Fig.~\ref{fig:s-lambda_nl3wr}), leading to an increase in the speed of sound. If the vector interaction is present in the quark phase, there is a competition of effects, and near to the pure-hadron boundary the drop in the speed of sound is less pronounced.

We further examine the polytropic index $\gamma$ along the three isentropic trajectories as a function of $\rho_B$ as shown in Fig.~\ref{fig:gamma-rho_nl3wr}. A horizontal dashed line marks the approximate criterion $\gamma < 1.75$ proposed in Ref.~\cite{Annala:2019puf} to signal the presence of quark matter at $T=0$. The $s/\rho_B = 0.5$ and $s/\rho_B = 2$ curves start above this threshold and decrease monotonically, exhibiting a rapid drop near the hadron–quark transition. By contrast, the $s/\rho_B = 5$ trajectory begins below rises to cross the threshold, and then drops again to values below 1.45; this behavior persists when $\zeta$ and/or $\alpha$ are non-zero.  Inside the coexistence region the trends mirror those seen for the sound velocity: the $s/\rho_B = 2$ and $s/\rho_B = 5$ curves decrease monotonically, whereas $s/\rho_B = 0.5$ develops a local maximum. As $\rho_B$ increases and the isentropes enter the pure-quark phase, $\gamma$ rises rapidly and subsequently drops, eventually approaching the conformal value $\gamma = 1$  at sufficiently high temperature and baryon density. We also find that $\gamma$ in the high-density quark phase is insensitive to the vector coupling (compare Figs.~\ref{fig:gamma-rho_nl3wr_a}, ~\ref{fig:gamma-rho_nl3wr_b} with Figs.~\ref{fig:gamma-rho_nl3wr_c}, ~\ref{fig:gamma-rho_nl3wr_d}, in agreement with the results of Ref.~\cite{Liu:2023gmq}.

\subsection{The role of hyperons}\label{wHyperons}

In the previous section, hyperonic degrees of freedom were neglected. At sufficiently large baryon densities, however, hyperons become energetically favorable and appear in chemical equilibrium with nucleons. In this section we study the impact of hyperons on the phase diagram and on thermodynamic observables, and we compare, qualitatively, those results with the hyperon-free case presented above. {For the hadronic sector we adopt the FSU2H parameterization, while for the quark phase we employ a smaller vector coupling of $\zeta=0.1$, in order to avoid having the speed of sound exceed the conformal limit too much, as discussed in the last section.

\begin{table*}[htpb]
    \centering
    \begin{tabular}{ c |c c c| c c c c || c c c c || c c c c || c c c c }
    \hline
        \multirow{2}{*}{Models} & \multirow{2}{*}{$\zeta$} & &\multirow{2}{*}{$\alpha$} &$m^*_u$ && $m^*_d$  && $m^*_u$ && $m^*_d$ && $m^*_u$ && $m^*_d$ && $m^*_u$ && $m^*_d$ &\\
        & & & & (MeV) && (MeV) && (MeV) && (MeV) && (MeV) && (MeV) && (MeV) && (MeV) & \\
    \hline
     \multicolumn{1}{c}{} & \multicolumn{3}{c}{} &\multicolumn{4}{c||}{$T = 10$ MeV} & \multicolumn{4}{c||}{$T = 80$ MeV} & \multicolumn{4}{c||}{$T = 100$ MeV} & \multicolumn{4}{c}{$T = 150$ MeV} \\
    \hline
    &0  & & 0    & 28.2 && 28.2  &&  21.2  && 21.2  && 21.6 && 21.6 && 75.4 && 75.4 & \\
   \multirow{2}{*}{\:NL3$\omega\rho$\:} &0.5 & & 0   & 17.4 && 17.4  && 15.9   && 15.9  && 16.7 && 16.7 && 49.7 && 49.7 & \\
    &0   & & 0.2  & 34.0 && 28.3  && 25.4   && 20.6 && 26.3 && 21.1 && 107.0 && 87.0 & \\
    &0.5 & & 0.2  & 19.3 && 16.8  && 17.6   && 15.3 && 18.7 && 16.1 && 61.3  && 51.5 & \\
    \hline
    &0   & & 0     & 21.5  && 21.5 && 16.5   && 16.5 && 17.2 && 17.2 && 53.6  && 53.6 & \\
    \multirow{2}{*}{\:FSU2H\:}& 0.1 & & 0  & 18.0  && 18.0 && 14.8   && 14.8 && 15.6 && 15.6 && 48.4  && 48.4 & \\
    &0   & & 0.2  & 26.3  && 21.7 && 20.2   && 16.2 && 21.0 && 16.8 && 67.8  && 56.3 & \\
    &0.1 & & 0.2 & 21.6  && 17.9 && 17.7   && 14.4 && 18.6 && 15.1 && 59.8  && 50.1 & \\   
    \hline
    \end{tabular}
    \caption{{The $u$, $d$ effective quarks masses at the onset of the mixed phase along the isotherms $T=$ 10, 80, 100, 150 MeV for the different scenarios and models discussed in the text.}}
    \label{tab:isotherms_quark_mass}
\end{table*}

The hadron–quark phase transition along several isotherms within the FSU2H-PNJL two-phase model is shown in Fig.~\ref{fig:isotherms_fsu2h}. Qualitatively, the results with and without hyperons (cf. Fig.~\ref{fig:isotherms_nl3wr}) are similar. {From Table~\ref{tab:isotherms_fsu2h}, however, one can see that the inclusion of hyperons systematically predicts the first-order transition to occur at higher pressures and baryon densities,  and the overall extent of the mixed-phase interval is slightly reduced}. For asymmetric matter, this reduction manifests itself as a steeper increase of pressure over the corresponding density range. Although the $\zeta = 0.1$ cases are not shown in Fig.~\ref{fig:isotherms_fsu2h}, their effect mimics that observed in Figs.~\ref{fig:isotherms_nl3wr_b} and~\ref{fig:isotherms_nl3wr_d} for the EoS NL3$\omega\rho$, namely a shift of the transition to higher density and pressure. In the present case, however, this shift is less pronounced owing to the smaller value of $\zeta$ used. The initial and final boundaries of the mixed phase in the $P-\rho_B$ plane for these configurations are nevertheless included in Table~\ref{tab:isotherms_fsu2h}.

{The EoS stiffness across the three phases can be studied using the speed of sound along each isentrope. Following this approach, Fig.~\ref{vs_rho_thermal} shows that despite FSU2H being the softer hadronic parameter set, it produces a comparatively stiffer EoS than NL3$\omega\rho$ within the mixed phase, that is a higher $c_s^2$ at equal $\rho_B$. Furthermore, the transition from hadron to quark matter occurs for a smaller density interval with FSU2H. One possible explanation is that the delayed onset of the mixed phase with the FSU2H EoS, results in larger energies and smaller quark masses (see Table~\ref{tab:isotherms_quark_mass}) when the mixed phase appears, which in turn drives the dissociation of hadrons into quarks more rapidly than with NL3$\omega\rho$. That more rapid conversion shortens the coexistence density range, yielding a stiffer mixed phase EoS. This behavior is illustrated in Fig.~\ref{vs_rho_thermal}, which shows the sound speed for the two parameterizations. For FSU2H the $T<150$ MeV isentropes enter the mixed phase with a relatively higher stiffness than for NL3$\omega\rho$; the rate at which stiffness changes is also higher for FSU2H. }

\begin{figure}[b!]
    \centering
    \includegraphics[width=\linewidth]{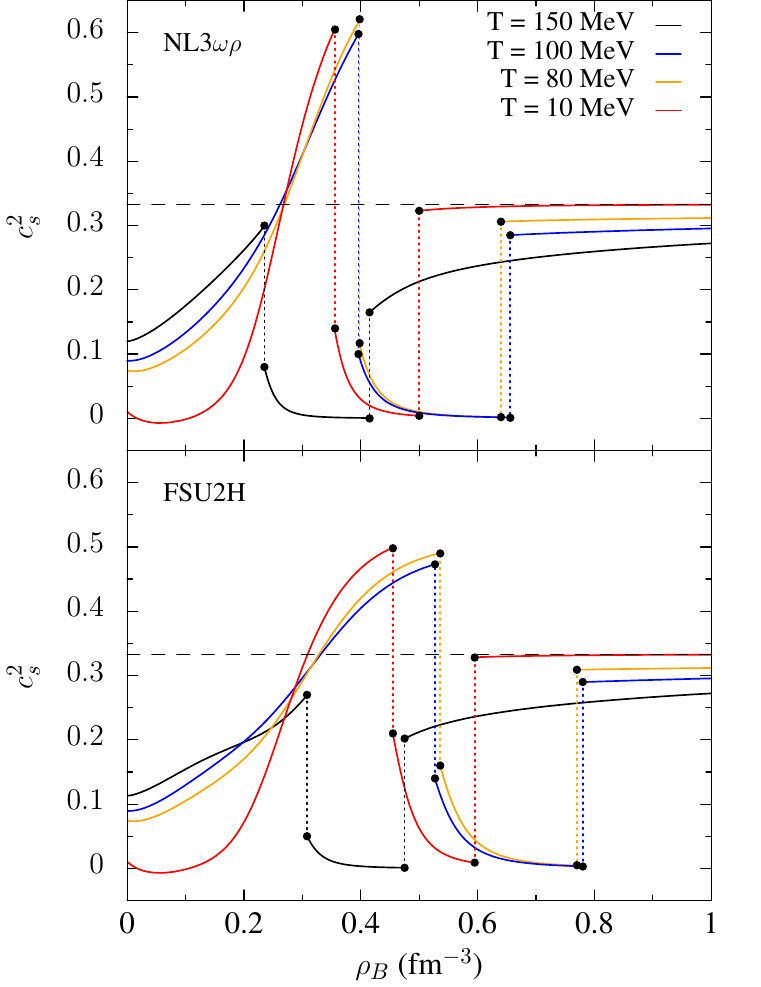}
    \caption{The squared speed of sound $c_s^2$ as a function of $\rho_B$ for $\zeta =0$ and $\alpha = 0.2$ at $T = 10$, 80, 100, 150 MeV using the hadronic NL3$\omega\rho$ (top) and FSU2H (bottom) models.}
    \label{vs_rho_thermal}
\end{figure}

\begin{figure*}[t!]
\centering
	\begin{subfigure}[b]{0.45\linewidth}
		\centering
		\includegraphics[width=\linewidth]{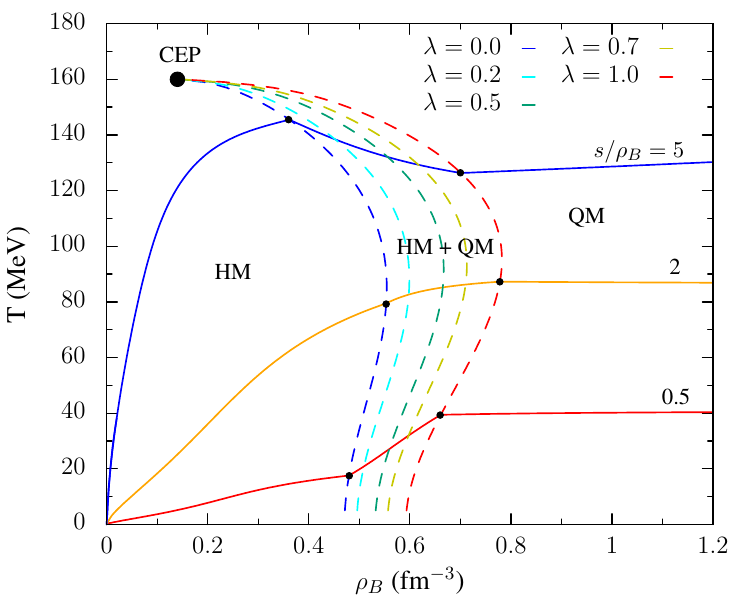}
		\caption{$\zeta = 0$ and $\alpha = 0$}
        \label{fig:T-rho_fsu2h_a}
	\end{subfigure}
    \hfil
	\begin{subfigure}[b]{0.45\linewidth}
		\centering
		\includegraphics[width=\textwidth]{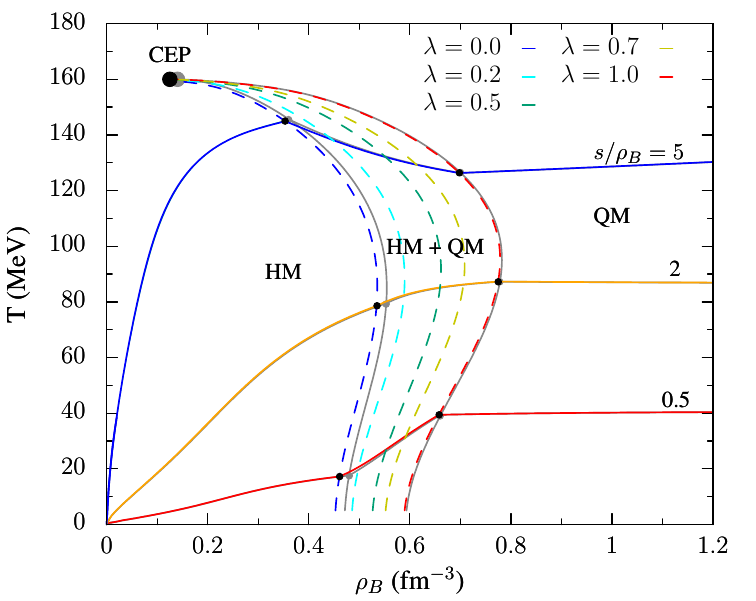}
		\caption{$\zeta = 0$ and $\alpha = 0.2$}
        \label{fig:T-rho_fsu2h_c}
	\end{subfigure}
	
    \caption{Phase diagram of the FSU2H-PNJL two-phase model in the $T-\rho_B$ plane for $\alpha = 0$ (left) and $\alpha = 0.2$ (right) matter with $\zeta = 0$. The colored dashed lines identify the coexistence region with different fractions of quark matter. The solid colored lines are isentropes for fixed $s/\rho_B =$ 0.5, 2, 5 ratios. For clarity, the $\lambda = 0$ and $\lambda = 1$ boundaries, together with the isentropes from panel~\ref{fig:T-rho_fsu2h_a}, are displayed in gray in panel~\ref{fig:T-rho_fsu2h_c} to highlight differences. }
	\label{fig:T-rho_fsu2h}
\end{figure*}

At temperatures close to the CEP, the delay in the onset of the mixed phase is reduced, reflecting both the presence of the CEP and its relative insensitivity to the choice of hadronic EoS (see Tables~\ref{tab:cep_nl3wr} and~\ref{tab:cep_fsu2h}). As a consequence, the onset density and pressure of the mixed phase become more similar for the two parameterizations, and density contributions to the stiffening of the mixed-phase EoS become less significant. In contrast to the behavior observed at lower temperatures, a stiffer hadronic EoS may therefore lead to a stiffer mixed-phase EoS, as seen for the $T=150$ MeV isotherm using the NL3$\omega\rho$ model. At this temperature, a global minimum of $c_s^2$ appears at the onset of the mixed phase among the selected isotherms. This behavior is consistent with predictions from effective models~\cite{Motta:2020cbr, He:2022kbc, He:2022yrk}, which indicate that the speed of sound reaches a global minimum at the CEP, although it does not vanish within the mean-field approximation.

The phase diagram of the FSU2H-PNJL two-phase model in the $T-\rho_B$ plane is shown in Fig.~\ref{fig:T-rho_fsu2h}. As in the NL3$\omega\rho$-PNJL case, isospin asymmetry has only a minor effect on the phase boundaries, resulting in a small earlier onset of the coexistence region at low temperature. Although not shown, quark vector interactions produce the same effect observed in Fig.~\ref{fig:T-rho_nl3wr}, namely a shift of the phase boundaries toward higher baryon densities and chemical potentials; however, this shift is less pronounced owing to the smaller coupling $\zeta = 0.1$. The CEP location is largely unchanged between the two models. For FSU2H the CEP lies at similar $T$ and $\mu_B$, with a slight shift to higher $\rho_B$ (see Table~\ref{tab:cep_fsu2h}). At low $T$, the hyperonic EoS noticeably shift the onset of the mixed phase, thereby disfavoring the emergence of quark matter at lower densities. This behavior follows directly from the softening of the hadronic EoS discussed above.

We also examine isentropic trajectories with fixed $s/\rho_B=$ 0.5, 2, 5 in the presence of hyperons (Fig.~\ref{fig:T-rho_fsu2h}). These isentropes exhibit an evolution similar to that obtained with the NL3$\omega\rho$ parameterization; however, their intersections with the phase boundaries are influenced by the onset of hyperons, which modify the hadronic composition and thermal response along each trajectory. Although all isentropes enter the coexistence region at higher baryon density, the corresponding onset temperature does not vary uniformly. Relative to NL3$\omega\rho$, the onset temperature in the FSU2H model remains nearly unchanged for $s/\rho_B=$ 0.5, increases for $s/\rho_B = 2$, and decreases for $s/\rho_B = 5$. This contrasting behavior can be traced to the different hyperon content along each trajectory. The evolution of the hyperon fractions with $\rho_B$ modifies the thermal and compositional properties of the hadronic phase and therefore changes where, and at which temperature, the isentropes intersect the coexistence region (see Fig.~\ref{fig_ratios_fsu2h}).

\begin{table}[h]
  
    \centering
    \begin{tabular}{c c c | c c c c c }
    \hline
        $\zeta$ && $\alpha$ & $T^{\rm CEP}$ (MeV) &&  $\rho_B^{\rm CEP}$ (fm$^{-3})$ && $\mu_B^{\rm CEP}$ (MeV) \\
    \hline
    0   && 0    & 160.0    && 0.140 && 516.0\\
    0.1 && 0    & 160.5    && 0.145 && 523.0 \\
    0   && 0.2  & 160.0    && 0.125 && 505.0 \\
    0.1 && 0.2  & 160.0    && 0.135 && 512.0 \\
    \hline
    \end{tabular}  
    \caption{Temperature, baryon density, and  baryon chemical potential for the CEP in the different scenarios considered within the FSU2H-PNJL model.}
    \label{tab:cep_fsu2h}
\end{table}


\begin{figure}[htpb]
    \centering
    \includegraphics[width=\linewidth]{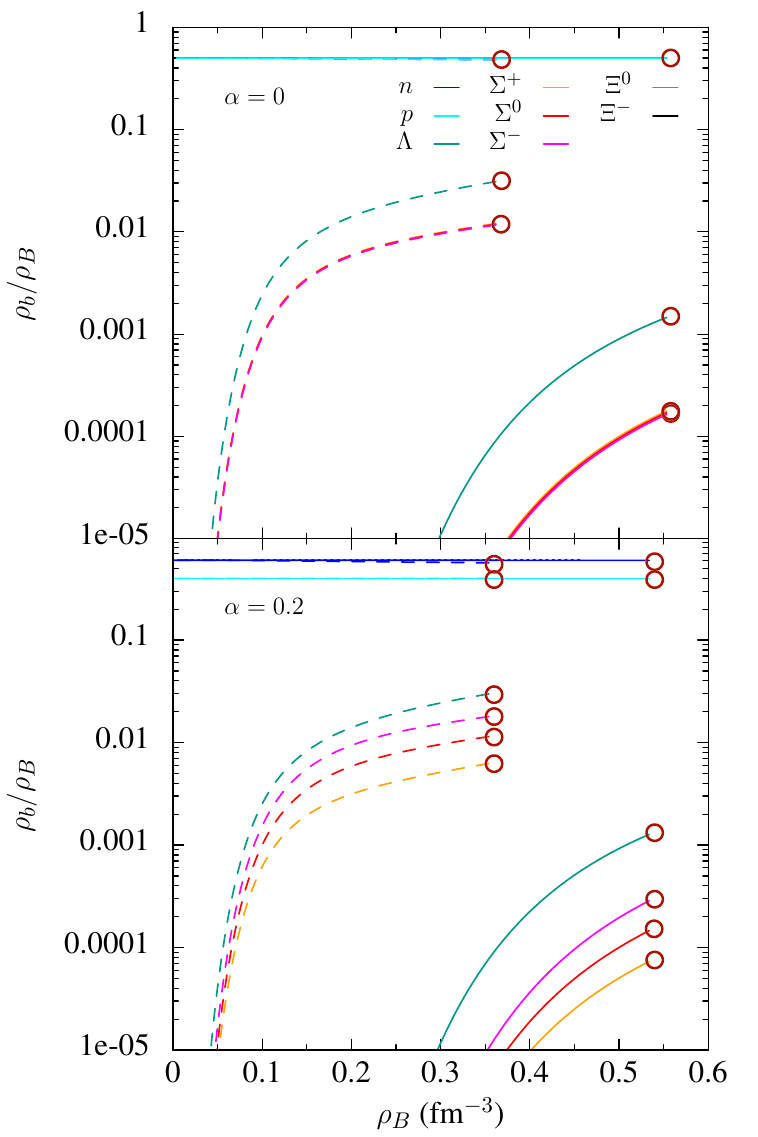}
    \caption{Baryon fractions as functions of the baryonic density for $\alpha = 0$ (top) and $\alpha = 0.2$ (bottom) in the FSU2H model with $\zeta = 0$. Full lines denote $s/\rho_B = 2$ isentropes, while dashed lines denote $s/\rho_B = 5$ isentropes. Open circles denote intersections with the $\lambda = 0$ boundary. The abundances of the $\Xi^0$, $\Xi^-$ baryons and  along the $s/\rho_B = 0.5$ trajectory are negligible and cannot be see in the plots. A log scale is used on the y axis. }
    \label{fig_ratios_fsu2h}
\end{figure}

\begin{figure*}[t!]
\centering
	\begin{subfigure}[b]{0.45\linewidth}
		\centering
		\includegraphics[width=\linewidth]{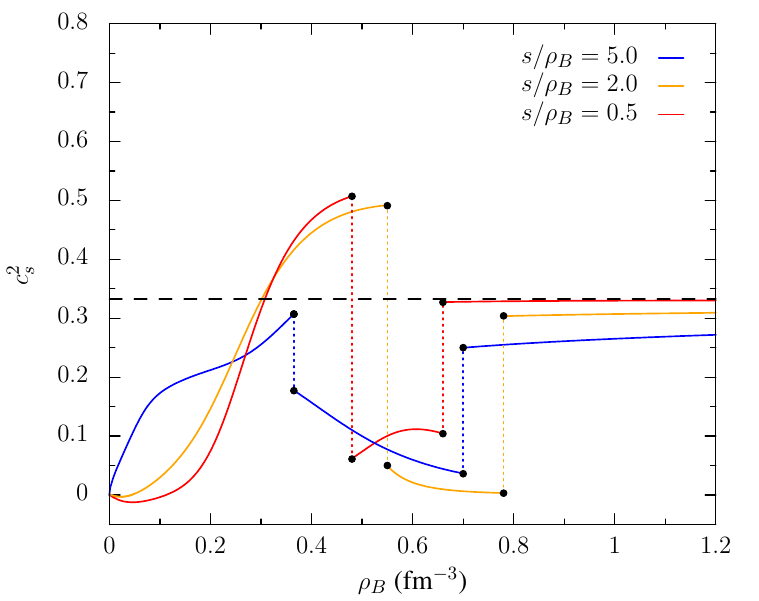}
		\caption{$\zeta = 0$ and $\alpha = 0$}
        \label{fig:vs-rho_fsu2h_a}
	\end{subfigure}
	\hfil
	\begin{subfigure}[b]{0.45\linewidth}
		\centering
		\includegraphics[width=\linewidth]{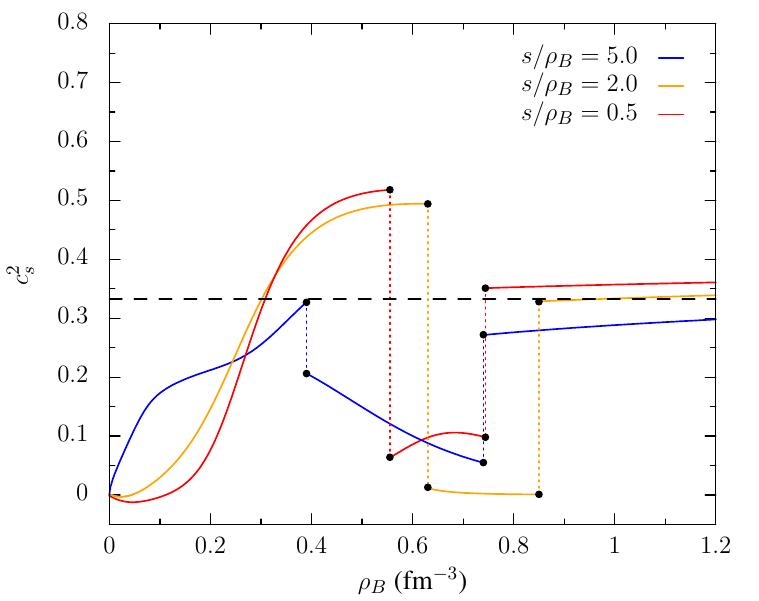}
		\caption{$\zeta = 0.1$ and $\alpha = 0$}
        \label{fig:vs-rho_fsu2h_b}
	\end{subfigure}
	\begin{subfigure}[b]{0.45\linewidth}
		\centering
		\includegraphics[width=\textwidth]{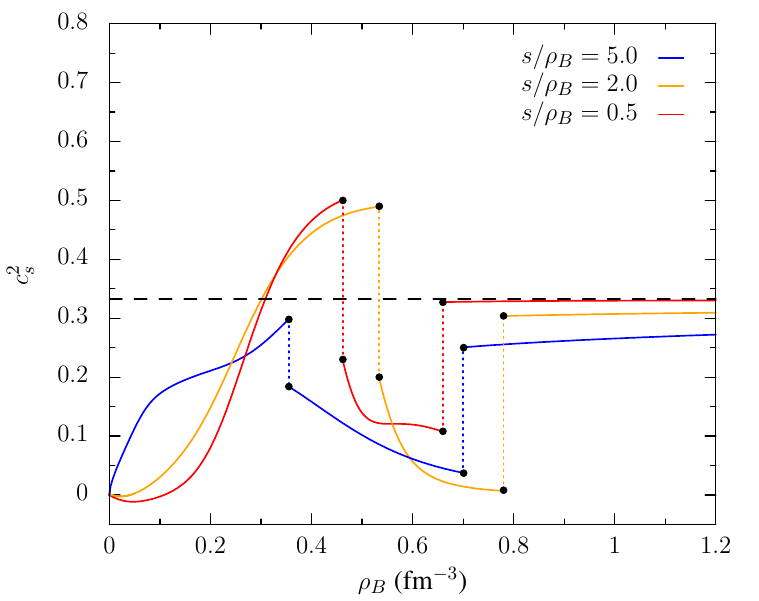}
		\caption{$\zeta = 0$ and $\alpha = 0.2$}
        \label{fig:vs-rho_fsu2h_c}
	\end{subfigure}
	\hfil
	\begin{subfigure}[b]{0.45\linewidth}
		\centering
		\includegraphics[width=\linewidth]{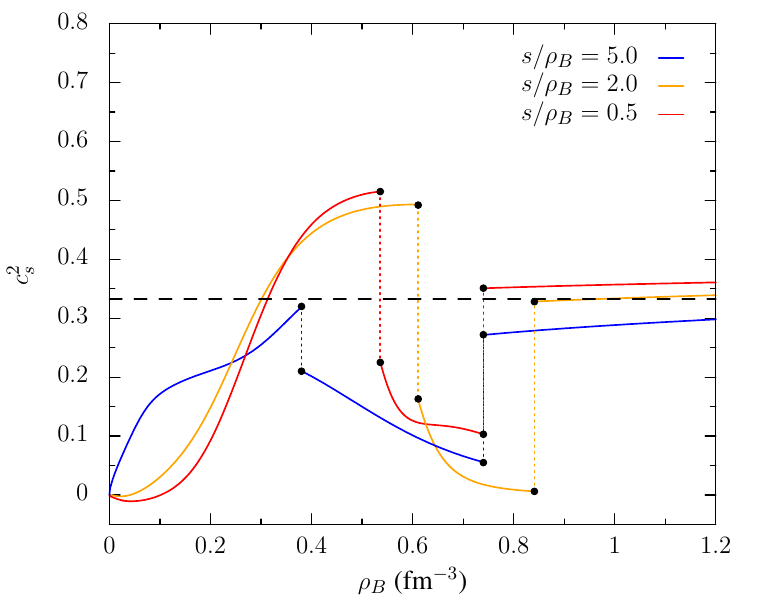}
		\caption{$\zeta = 0.1$ and $\alpha = 0.2$}
        \label{fig:vs-rho_fsu2h_d}
	\end{subfigure}
    
    \caption{Sound velocity squared $c_s ^2$ as a function of the baryon density along the isentropes $s/\rho_B = 0.5$, 2, 5 for $\zeta = 0$ (left) and $\zeta = 0.1$ (right), and for $\alpha = 0$ (top) and $\alpha = 0.2$ (bottom) within the FSU2H-PNJL two-model approach. The horizontal dashed line $c_s^2 = 1/3$ indicates the high-density conformal limit.}
	\label{fig:vs-rho_fsu2h}
\end{figure*}

\begin{figure*}[t!]
\centering
	\begin{subfigure}[b]{0.45\linewidth}
		\centering
		\includegraphics[width=\linewidth]{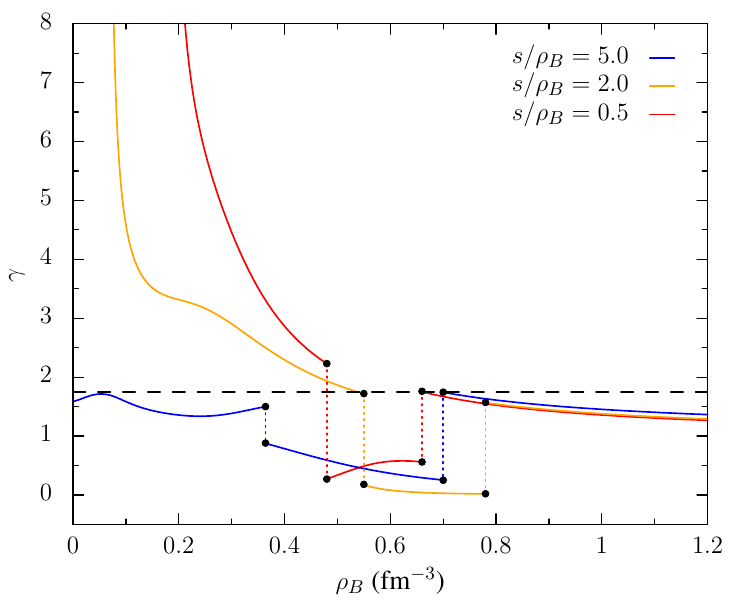}
		\caption{$\zeta = 0$ and $\alpha = 0$}
        \label{fig:gamma_fsu2h_a}
	\end{subfigure}
	\hfil
	\begin{subfigure}[b]{0.45\linewidth}
		\centering
		\includegraphics[width=\linewidth]{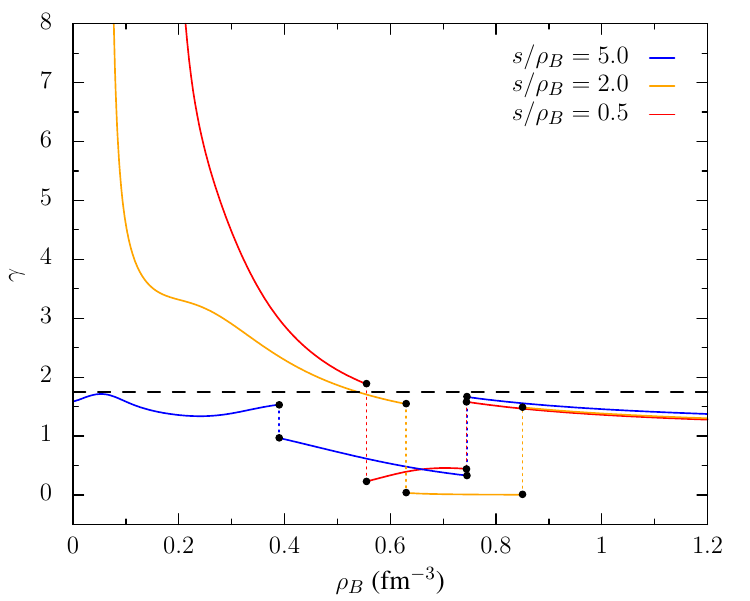}
		\caption{$\zeta = 0.1$ and $\alpha = 0$}
        \label{fig:gamma_fsu2h_b}
	\end{subfigure}
	\begin{subfigure}[b]{0.45\linewidth}
		\centering
		\includegraphics[width=\textwidth]{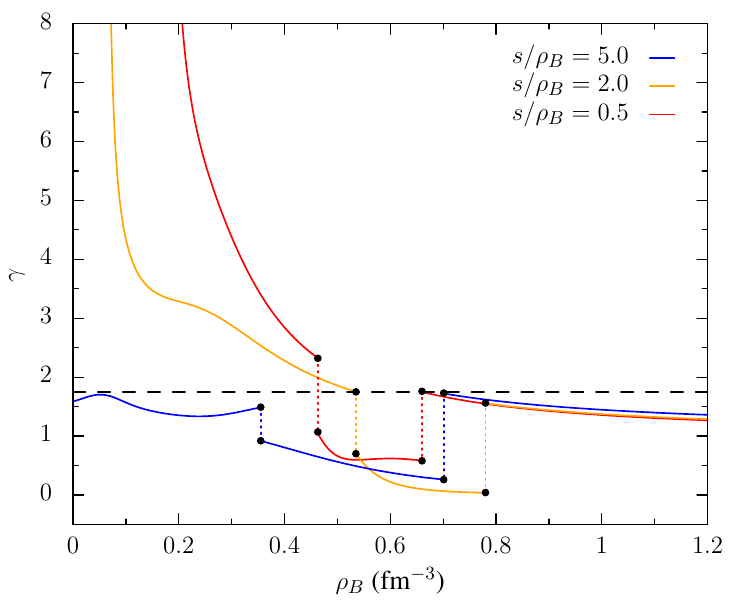}
		\caption{$\zeta = 0$ and $\alpha = 0.2$}
        \label{fig:gamma_fsu2h_c}
	\end{subfigure}
	\hfil
	\begin{subfigure}[b]{0.45\linewidth}
		\centering
		\includegraphics[width=\linewidth]{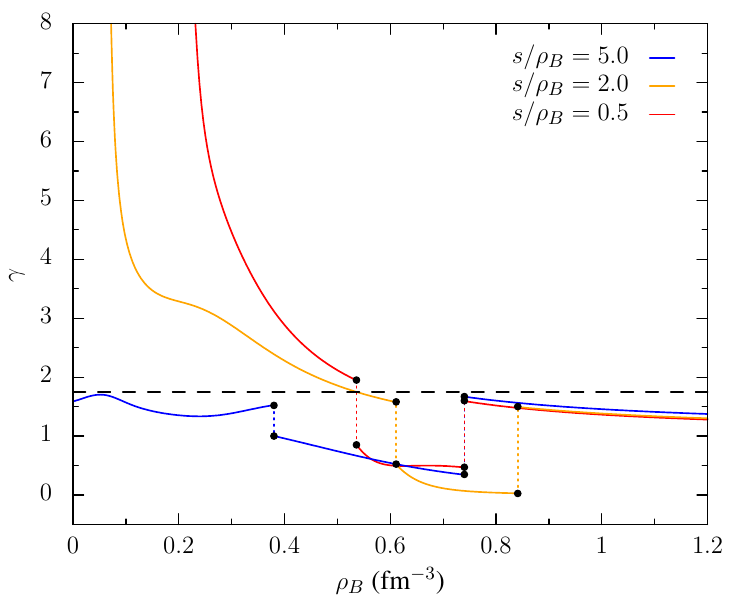}
		\caption{$\zeta = 0.1$ and $\alpha = 0.2$}
        \label{fig:gamma_fsu2h_d}
	\end{subfigure}
    
    \caption{Polytropic index $\gamma$ as a function of the baryon density along the isentropes $s/\rho_B =0.5$, 2, 5 for $\zeta = 0$ (left) and $\zeta = 0.1$ (right), and for $\alpha = 0$ (top) and $\alpha = 0.2$ (bottom) within the FSU2H-PNJL two-model approach. The horizontal dotted line $\gamma = 1.75$ serves as an approximate reference value to distinguish hadronic from quark degrees of freedom.}
	\label{fig:gamma-rho_fsu2h}
\end{figure*}

Figure~\ref{fig_ratios_fsu2h} displays the hyperon number density fractions along the isentropic trajectories. The upper panel shows the isospin-symmetric case $(\alpha = 0)$, while the lower panel corresponds to the isospin-asymmetric case $(\alpha=0.2)$. Isospin asymmetry modifies the relative populations within the isospin doublet \((n,p)\) and the isospin triplet \((\Sigma^-,\Sigma^0,\Sigma^+)\), producing noticeable changes in their abundances. Solid lines denote the intermediate-entropy isentrope \(s/\rho_B=2\) and dashed lines denote the high-entropy isentrope $s/\rho_B=5$. The \(\Xi^0\) and \(\Xi^-\)  fractions, as well as the abundances along the low-entropy trajectory $s/\rho_B =0.5$, are negligible.

The splitting of the \(\Sigma\)-baryon abundances is governed by the coupling to the \(\rho\)-meson. In isospin-symmetric matter the \(\rho\)-meson mean field vanishes and no isospin splitting occurs. For \(\alpha\neq 0\), however,  $\rho$-field is finite and the isovector coupling term $g_{\rho b} \gamma_\mu \boldsymbol{\tau}_b\cdot \boldsymbol{\rho}^\mu$ in Eq.~\eqref{lb_nl3wr} contributes to the effective chemical potential. Because $\tau^3_{\Sigma^\pm}=\pm 1$, the $\rho$-field lowers the effective chemical potential of \(\Sigma^+\) and raises that of \(\Sigma^-\), reducing the \(\Sigma^+\) density fraction while enhancing  \(\Sigma^-\). The same mechanism produces the relative-abundance splitting in the $(n,p)$ doublet.

{In both panels hyperons appear at lower densities along the \(s/\rho_B=5\) isentrope because the temperature increases more rapidly along that trajectory, enhancing thermal excitations. Consequently, when the \(s/\rho_B=5\) isentrope reaches the mixed phase the hyperon fractions are larger by approximately one order of magnitude than along the \(s/\rho_B=2\) isentrope. This earlier and more abundant hyperon population causes an earlier softening of the hadronic EoS, which in turn reduces the temperature along the high-entropy isentrope as it approaches the phase boundary. By contrast, the \(s/\rho_B=0.5\) and $s/\rho_B = 2$ isentropes experience a milder temperature rise and smaller thermal contributions. For $s/\rho_B = 0.5$ hyperons are nearly absent, whereas for $s/\rho_B = 2$ they become relevant only above \(\rho_B\gtrsim 0.3\ \mathrm{fm}^{-3}\). This situation therefore provides a wider density interval over which the isentrope can heat before additional degrees of freedom emerge. Because the onset of the mixed phase occurs at higher $\rho_B$ for FSU2H-PNJL, the low- and intermediate-entropy isentropes intersect the coexistence region at higher temperatures than in the NL3$\omega\rho$-PNJL model.}

As in the NL3$\omega\rho$-PNJL case, the onset of the $s$-quark degrees of freedom in the FSU2H-PNJL model lies outside the mixed-phase density interval shown in Fig.~\ref{fig:T-rho_fsu2h}. The $s$-quark crossover occurs at $\rho_B \approx 1.4$ fm$^{-3}$, so its influence on the isentropic trajectories is only apparent for those that probe such large baryon densities.

Similar to Fig.~\ref{fig:vs-rho_nl3wr} and Fig.~\ref{fig:gamma-rho_nl3wr} for NL3$\omega\rho$, Fig.~\ref{fig:vs-rho_fsu2h} and Fig.~\ref{fig:gamma-rho_fsu2h} show the speed of sound and the polytropic index for the FSU2H model. Several general features persist: a liquid–gas instability at very low density (signaled by $c_s^2< 0$ along the $s/\rho_B = 0.5$ and 2 isentropes) and steep discontinuities at the coexistence-region boundaries; for larger $\rho_B$ we always find $c_s^2 > 0$. However, when hyperons are included the magnitude of these discontinuities and the changes on the sound speed along each trajectory depend on the chosen $s/\rho_B$ ratio. In the hadronic phase, the maximum  $c_s^2$ for the low- and intermediate-entropy curves is substantially reduced at the onset of the mixed phase, despite the larger onset baryon density. By contrast, for $s/\rho_B = 5$ the larger $\rho_B$ coincides with a larger peak of $c_s^2$. This behavior results from the interplay between the softening (induced by hyperons) of the hadronic EoS and thermal effects. For the $s/\rho_B = 0.5$ and 2 isentropes, hyperon-driven softening dominates, and the larger onset $\rho_B$ is insufficient to compensate the reduction of the pressure-to-energy-density ratio. Indeed, whereas for the NL3$\omega\rho$ EoS (Fig.~\ref{fig:vs-rho_nl3wr}) the maximum for $s/\rho_B =0.5$ lies below that for $s/\rho_B =2$, for FSU2H (Fig.~\ref{fig:vs-rho_fsu2h}) this ordering is inverted, reflecting the more abundant hyperon population along the $s/\rho_B = 2$ trajectory. For $s/\rho_B = 5$, however, the rapid temperature rise dominates, and thermal excitations largely offset the pressure reduction cause by the appearing of hyperons, so the higher onset $\rho_B$ produces a larger maximum $c_s^2$.

At the $\lambda = 0$ boundary, the magnitude of the steep discontinuities is reduced for the $s/\rho_B = 0.5$ and 2 isentropes but increases for $s/\rho_B = 5$, relative to NL3$\omega\rho$. In Fig.~\ref{fig:vs-rho_fsu2h_a}, the mixed-phase EoS evolves similarly as for NL3$\omega\rho$; in particular, a peak value in $c_s^2$ persists along the $s/\rho_B = 0.5$ trajectory. {Although the intermediate- and high-entropy curves enter the mixed phase with a lower $c_s^2$, the mixed-phase EoS is stiffer (higher $c_s^2$) than for NL3$\omega\rho$ over the same baryon density interval. By contrast, the low-entropy trajectory enters the mixed phase with a higher $c_s^2$ because of the larger onset density; along this trajectory the mixed-phase EoS is initially softer (lower $c_s^2$) but becomes stiffer (higher $c_s^2$) than NL3$\omega\rho$ at sufficiently high density.} The effects of isospin asymmetry shown in Fig.~\ref{fig:vs-rho_fsu2h_c} are identical to those  describe for NL3$\omega\rho$: a local minimum forms in the $s/\rho_B= 0.5$ curve, and the discontinuities at the onset density are reduced, particularly for the low- and intermediate-entropy isentropes. {Here, although the onset value of  $c_s^2$ is again higher for $s/\rho_B = 0.5$ and lower for $s/\rho_B =$ 2 and 5, the mixed-phase EoS for FSU2H remains stiffer than that of NL3$\omega\rho$ along the same trajectory and over the same density range, mostly owning to the larger onset densities and pressures due to hyperons.}
The addition of quark vector interactions (Figs.~\ref{fig:vs-rho_fsu2h_b} and~\ref{fig:vs-rho_fsu2h_d}) produces effects similar to those described before; for the smaller coupling shown ($\zeta = 0.1$) the peak seen in Figs.~\ref{fig:vs-rho_nl3wr_b} and~\ref{fig:vs-rho_nl3wr_d} is absent. We observe that even for the smaller vector coupling $\zeta = 0.1$ the low- and intermediate-entropy trajectories exceed the $c_s^2 =1/3$ limit in the quark phase already for $\rho_B \gtrsim 0.7$, within the model’s validity range. {Indeed, it is known that a drawback of the including a finite vector coupling is that the EoS remains stiff at higher densities so that the conformal limit, observed by perturbative QCD (pQCD), cannot be attained. One possible way to circumvent this problem is to consider a density-dependent vector coupling, $G_V(\rho_B)$, able to interpolate between a regime where repulsion is high (the EoS is stiff) and a regime where repulsion is low (the EoS is soft), while conciliating neutron stars' constraints at lower densities and pQCD predictions at asymptotically high baryonic densities. The simple \textit{ansatz} proposed in Ref.~\cite{Pinto:2022lkv} for $G_V(\rho_B)$ in the three-flavor NJL model approaches the conformal limit at high density, while supporting the claim in Ref.~\cite{Fujimoto:2022ohj} that the presence of a nonconformal peak in $c_s^2$ is not necessarily in tension with the trace anomaly being positive for all densities -- a nontrivial results cannot be simultaneously obtained when $G_V$ vanishes or has a fixed value~\cite{Pinto:2022lkv}. Such study is beyond the scope of the present work and will be saved for future investigation.}

Figure~\ref{fig:gamma-rho_fsu2h} displays the polytropic index $\gamma$ for the FSU2H model. The evolution of $\gamma$ along each isentrope closely resembles that in Fig.~\ref{fig:gamma-rho_nl3wr}. {As for NL3$\omega\rho$, the approximate criterion $\gamma \leq 1.75$ (at $T=0$) proposed in Ref.~\cite{Annala:2017llu} remains a useful indicator for distinguishing hadronic from quark matter along the $s/\rho_B = 0.5$ and $2$ trajectories. However, for $s/\rho_B = 5$, i.e., at higher $T$, this criterion loses its discriminating power, as $\gamma$ remains mostly below 1.75 across all three phases. In this regime, a more restrictive interval, $\gamma \lesssim 1-1.4$, appears to provide a more appropriate reference for identifying  the possible onset of the quark matter at finite $T$. Within the hadron-quark mixed phase $\gamma$ is generally below the conformal limit; at still larger densities (in the quark phase) $\gamma$ appears to approaches the conformal value from above.}

\section{SUMMARY}\label{sec:summary}

In the work, we investigated the hadron–quark phase transition using a two-phase construction based on a three-flavor PNJL model for the quark sector and two hadronic EOS of different stiffness at intermediate densities. We analyze the combined effects of isospin asymmetry, quark vector interactions, and the thermal and density dependence of the EoS on the phase boundaries, the speed of sound and the polytropic index along isentropic and isothermal trajectories, including in the vicinity of the CEP. We also examined the role of hyperonic degrees of freedom and their impact on the phase diagram and along each trajectory.

Increasing isospin asymmetry lowers the onset density and temperature of the phase transition, suggesting that neutron-rich systems may probe the transition more efficiently in HICs. A harder hadronic EoS (NL3$\omega\rho$) shifts the transition to lower densities and pressures, favoring the earlier onset of the mixed phase, although this effect remains small for asymmetries relevant to current HIC experiments. In contrast, quark vector interactions significantly displace the phase diagram toward higher baryon density and chemical potential, and broaden the mixed-phase region in the low temperature region.

A CEP is present in all scenarios considered. Its position is relatively stable but shifts systematically toward higher baryon densities when quark vector interactions are included, particularly in asymmetric matter. This contrasts with results from single-model approaches, where the CEP is highly sensitive to model parameters~\cite{Costa:2008gr, Costa:2009ae} and may disappear for strong vector couplings~\cite{Costa:2010zw, Ferreira:2018sun, Costa:2020dgc}. For the softer hadronic EoS with hyperons (FSU2H), the mixed phase is shifted to higher densities and reduced in extent, yet the CEP location remains close to that obtained with NL3$\omega\rho$.

The thermal evolution within the mixed phase depends strongly on the entropy per baryon. Low- and intermediate-entropy trajectories show monotonic heating as the system evolves from hadronic to quark matter, whereas high-entropy trajectories near the CEP exhibit cooling. Quark vector interactions weaken the heating effect and enhance cooling. These behaviors reflect differences in the entropy content of the coexisting phases and the interplay of density effects, thermal excitations, and effective masses. When hyperons are included, changes in their population further influence the heating and cooling behavior within the mixed phase, as well as the onset temperature, despite the systematic shift of the onset density to higher values.

The speed of sound and the polytropic index display clear signatures of the phase transition. Away from the CEP, a softer hadronic EoS leads to a relatively stiffer mixed-phase EoS due to its higher onset density and energy density. Near the CEP, however, a pronounced minimum in $c_s^2$ develops at the transition onset. Both $c_s^2$ and the $\gamma$ exhibit discontinuities at the phase boundary, reflecting the abrupt change in degrees of freedom, which gradually smooth out in the quark phase. The impact of quark vector interactions depends on $s/\rho_B$: they enhance the discontinuity at low entropy but reduce it at high entropy. Such differences may influence the dynamical evolution of the fireball in HICs, particularly during the emergence of quark degrees of freedom, and could affect the experimental search for signals of the hadron–quark phase transition and the CEP. The evolution of $c_s^2$ within the mixed phase shows entropy-dependent peak and dip structures also dependent on $s/\rho_B$, particularly in asymmetric matter. The criterion  $\gamma\leq 1.75$ remains useful for identifying quark matter along low- and intermediate-entropy trajectories (lower $T$). However, near to the CEP (higher $T$), $\gamma$ remains mostly below 1.75 across all three phases, and a narrower interval ($\gamma \lesssim 1-1.4$) seems provides a more reliable reference for identifying the possible onset of the quark matter at finite $T$.

\begin{acknowledgments}
This work was partially supported by national funds from FCT (Fundação para a Ciência e a Tecnologia, I.P, Portugal) under project UID/04564/2025 identified by DOI 10.54499/UIDB/04564/2025.  
\end{acknowledgments}

\appendix
\section{FORMALISM}\label{sec:app_formalism}
\subsection{The PNJL model}

The thermodynamical potential for the three-flavor PNJL model reads	
\begin{align}
    \Omega(T, \mu_i) &= G_{S}\left(\sigma_u^2+ \sigma_d^2 + \sigma_s^2\right) 
    + 4 G_D\sigma_u\sigma_d\sigma_u \notag\\
    & +G_V \left( \rho_u^2 + \rho_d^2 + \rho_s^2 \right) - {\cal U}\left(\Phi,\bar{\Phi},T\right)\notag\\
    & - 2T\sum_{i=u,d,s}\int\frac{\mathrm{d}^3p}{\left(2\pi\right)^3}\left(N_c\beta E_i + z^+_\Phi(E_i) + z^-_\Phi(E_i)\right),
\end{align}
where  $E_{i}=\sqrt{{p}_i^{2}+{m^*_{i}}^{2}}$ is the quasiparticle energy of the quark $i$, and $z^{\pm}_\Phi$ represent the following partition 
function densities,
\begin{equation}
\begin{split}
	z^+(E_i) &=	\ln\left(1 + N_c\bar\Phi e^{-\beta E^{+}} + N_c\Phi e^{-2\beta E^{+}} + e^{-3\beta E^{+}}\right),\\
		z^-(E_i) &=  \ln\left(1 + N_c\Phi e^{-\beta E^{-}_i} + N_c\bar\Phi e^{-2\beta E^{-}_i} + e^{-3\beta E^{-}}\right),
\end{split}
\end{equation}
where $E_i^{(\pm)}=E_i\mp \muR_i$ with the upper (lower) sign 
applying for fermions (antifermions), and $\beta=1/T$.
The quark effective chemical potentials are given by
$$  
{\muR_i}=\mu_i-4G_V\rho_i.
$$
The $i$ quark number density is determined by $\rho_i = - (\partial\Omega/\partial\mu_i)$ and reads
\begin{equation}
\rho_i=-2N_c\int \frac{{\rm d} ^3p}{(2\pi)^3}\left(\F^{+}_i(E_i)-\F^{-}_i(E_i) \right).
\end{equation}
The modified Fermi-Dirac distribution functions $\F^+_i(E_i)$ and 
$\F^{-}_i(E_i)$ are given by

\begin{equation}
	\label{pnjl_occupation_numbers}
	\begin{split}
		\F^+_i(E_i) &= \frac{\frac{3}{N_c}e^{-3\beta E_i^+} + \Phibar e^{-\beta E_i^+} + 2\Phi e^{-2\beta E_i^+}}{1 + e^{-3\beta E_i^+} + N_c\Phibar e^{-\beta E_i^+} + N_c\Phi e^{-2\beta E_i^+}},\\[6pt]
	\F^-_i(E_i) &= \frac{\frac{3}{N_c}e^{-3\beta E_i^-} + \Phi e^{-\beta E_i^-} + 2\Phibar e^{-2\beta E_i^-}}{1 + e^{-3\beta E_i^-} + N_c\Phi e^{-\beta E_i^-} + N_c\Phibar e^{-2\beta E_i^-}}.
	\end{split}
\end{equation}
In the mean-field approximation, the values of the quark condensates are given by 
\begin{equation}
\sigma_i =-2N_c\int\frac{\mathrm{d}^3p}{\left(2\pi\right)^3}
\frac{m^*_i}{E_i}\left(1-\F^{+}_i(E_i)-\F^{-}_i(E_i)\right),
\end{equation}
which satisfy the following gap equations:
\begin{align}
m^*_i&=m_i-2G_S\sigma_i-2G_D\sigma_j\sigma_k,
\end{align}
with $i\neq j\neq k \in \{u, d,s\}$.
The values of $\Phi$ and $\bar{\Phi}$ are the solutions of 
\begin{equation}
	\begin{split}
		-\frac{a(T)}{2}\Phibar &=
	  \frac{6b(T)(\Phibar - 2\Phi^2 + \Phibar^2 \Phi)}{1 - 6\Phibar\Phi + 4(\Phibar^3 + \Phi) - 3(\Phibar\Phi)^2}\\
	   &+ \frac{2N_c}{T^3}\sum_{f=u,d,s}\int \frac{\diff^3p}{(2\pi)^3}\left(\frac{e^{-\beta E_i^- }}{e^{z^-_i(E_i)}} + \frac{e^{-2\beta E_i^+}}{e^{z_i^+(E_i)}}\right),\\[6pt]
		-\frac{a(T)}{2}\Phi &=
		\frac{6b(T)(\Phi - 2\Phibar^2 + \Phibar \Phi^2)}{1 - 6\Phibar\Phi + 4(\Phibar^3 + \Phi) - 3(\Phibar\Phi)^2}\\
		&+ \frac{2N_c}{T^3}\sum_{f=u,d,s}\int \frac{\diff^3p}{(2\pi)^3}\left(\frac{e^{-\beta E^+_i}}{e^{z^+_i(E_i)}} + \frac{e^{-2\beta E^-_i}}{e^{z^-_i(E_i)}}\right).
	\end{split}\label{polyakov_loop_mfa_eqs}
\end{equation}
The thermodynamical quantities can be determined via the thermodynamical potential 
(see~\cite{Costa:2010zw}).
The pressure is given by
\begin{equation}
     P(T,\mu_i)=-\Omega(T,\mu_i),\label{pressure_thermPot}
\end{equation}
the density of the $i$ quark, $\rho_i$, by
\begin{equation}
\rho(T,\mu_i)\,= \left( \frac{\partial P}{\partial {\mu_i}}\right)_{T},\label{rho_pressure}
\end{equation}
while the the entropy density, $s$, is given by
\begin{equation}
s(T,\mu_i)\,= \left( \frac{\partial P}{\partial T}\right)_{\mu_i}\label{entropy_pressure}.
\end{equation}
The energy density, $ \mathcal{E}$, comes from the following fundamental relation of
thermodynamics
\begin{equation}
     \mathcal{E} (T, \mu_i)=Ts(T,\mu_i) - P(T,\mu_i)+ \sum_{i=u,d,s}\mu_i\rho_i\label{therm_energy_relation}.
\end{equation}

In this work, for the vector channel, we adopted the interaction of Eq.~\eqref{p1} which generates a repulsive contribution proportional to the quark densities and leads to shifted effective chemical potentials for each flavor. In an isospin-symmetric case, the resulting thermodynamic structure is qualitatively similar to that obtained from a purely isoscalar interaction of the form $((\bar{\psi}\gamma_\mu\psi)^2)$. However, in asymmetric matter the distinction becomes more relevant. A purely isoscalar interaction depends only on the total quark density and contributes to the thermodynamic potential through a term proportional to the square of the total baryon density. On the other hand, a flavor-dependent interaction introduces separate contributions associated with each flavor density and therefore modifies differently the effective chemical potentials of the $u$ and $d$ quarks in asymmetric matter. Complete studies can be found in Refs. \cite{Masuda:2012ed,Ferreira:2021osk,Ali:2024owl}.

\subsection{Hadronic models}

Within the relativistic mean-field approach, the thermodynamical potential for the NL3$\omega\rho$ Lagrangian density is 
\begin{align}
\label{nl3wr_therm_potential}
		\Omega(T, \mu_b) &=  \frac{1}{2}m_\sigma^2\sigma^2 +\frac{\kappa}{3!}(g_{\sigma N}\sigma)^3 + \frac{\lambda}{4!}(g_{\sigma N}\sigma)^4\notag\\
		&- \frac{1}{2}m^2_\omega \omega_0^2  - \frac{\xi}{4!}(g_{\omega N}\omega_0)^4\notag\\
		&- \frac{1}{2}m^2_\rho(\rho_0^3)^2 - \Lambda_\omega g^2_{\rho N}g^2_{\omega N}(\rho_0^3)^2 (\omega_0)^2\notag\\
		&-2T\sum_{b} \int \frac{{\rm d}^3 p}{(2\pi)^3} \left[ \ln\left(1 + e^{-\beta\left(E_b + \mu_{ b}^*\right)}\right)\right.\notag\\
        &+\left. \ln\left(1 + e^{-\beta\left(E_b - \mu_{ b}^*\right)}\right)\right],
\end{align}
with $E_b = \sqrt{{p}^2_b + {m^*_b}^2}$, the effective chemical potential of baryon $b$ 
\begin{equation}
    \mu_b^* = \mu_b - g_{\omega b}\omega_0 - g_{\rho b}\tau_{3b}\rho_0^3 \label{baryon_chem_pot},
\end{equation}
and the baryon effective mass 
\begin{equation}
    m_b^* = m_b - g_{\sigma b}\sigma.
\end{equation}
The equations of motion for the meson fields in the mean-field approximation for the uniform medium are
	\begin{align}
		&m_\sigma^2 \sigma + \frac{\kappa}{2}g^3_{\sigma N}\sigma^2 + \frac{\lambda}{3!}g^4_{\sigma N}\sigma^3 = \sum_{b}g_{\sigma b}\rho_b^s,\label{eq.motion.H.a}\\ 
		&m_\omega^2\omega_0 + \frac{\xi}{3!}g^4_{\omega N}\omega_0^3 + 2\Lambda_\omega g^2_{\rho N}g^2_{\omega N} (\rho^3_0)^2\omega_0 = \sum_{b}g_{\omega b}\rho_b,\label{eq.motion.H.b}\\ 
		&m_{\rho}^2 \rho^3_0 +  2\Lambda_\omega g^2_{\rho N}g^2_{\omega N} \rho^3_0(\omega_0)^2 = -\sum_{b}g_{\rho b}\tau_{3b} \rho_b.\label{eq.motion.H.c}
	\end{align}
Here, 
\begin{equation}
    \rho_b^s = 2\int \frac{{\rm d}^3 p}{(2\pi)^3}\frac{m^*_b}{E_b}\left(f_b^+(E_b) + f_b^-(E_b)\right),
\end{equation}
is the baryon scalar density of particle $b$ , and the respective baryon density is
\begin{equation}
   \rho_b =  2\int\frac{{\rm d}^3 p}{(2\pi)^3}\left(f^+_b(E_b) - f^-_b(E_b)\right),
\end{equation}
where $f^\pm_b$ is the Fermi distribution for the baryon ($+$) and antibaryon ($-$) $b$, given by
\begin{equation}
	f_b^\pm(E_b) = \frac{1}{e^{\beta(E_b \mp \mu^*_b)} + 1} \label{ocupation numbers}.
\end{equation}

For the FSU2H parametrization, the thermodynamical potential~\eqref{nl3wr_therm_potential} is extended to include the $\phi$-field meson by adding the mass term
\begin{equation}
    \Omega(T, \mu_b) \to \Omega(T,\mu_b) - \frac{1}{2}m_\phi^2\phi_0^2,
\end{equation}
and a $\phi$-baryon coupling term to the effective chemical potential~\eqref{baryon_chem_pot}
\begin{equation}
    \mu_b^* \to \mu^*_b - g_{\phi b}\phi_0.
\end{equation}
The equation of motion for the $\phi$-meson mean field,
\begin{equation}
    m_\phi^2 \phi_0 = \sum_{b}g_{\phi b}\rho_b,\label{eq.motion.H.d} 
\end{equation}
is included together with Eqs.~\eqref{eq.motion.H.a}-\eqref{eq.motion.H.c}. The bulk thermodynamic properties of the system are then obtained by applying the relations given in Eqs.~\eqref{pressure_thermPot}–\eqref{therm_energy_relation}.

\section{HADRONIC PARAMETERIZATIONS}\label{sec:app_hadron_params}
The hadronic phase is described by a relativistic mean-field Lagrangian with two parameter sets -- NL3$\omega\rho$~\cite{Lalazissis:1996rd,Carriere:2002bx} and FSU2H~\cite{Tolos:2016hhl,Tolos:2017lgv} -- in order to compare scenarios that exclude and include hyperon degrees of freedom. The model couplings are determined from a multi-parameter fit to finite-nuclei properties and astrophysical observations. The coupling constants and saturation properties for both parameterizations are reported in Table~\ref{tab.Params_nlw_fus2h}.

The hyperon-vector couplings are obtained by assuming SU(3) flavor symmetry together with ideal $\omega-\phi$ mixing, which yields the relative strengths~\cite{Weissenborn:2011ut}
\begin{equation}
	\begin{gathered}
		g_{\omega \Lambda} : g_{\omega\Sigma}:g_{\omega\Xi}:g_{\omega N} = \frac{2}{3} : \frac{2}{3} : \frac{1}{3} : 1,\\
		g_{\phi\Lambda} : g_{\phi\Sigma} : g_{\phi\Xi} : g_{\omega N} = -\frac{\sqrt{2}}{3} : -\frac{\sqrt{2}}{3} : -\frac{2\sqrt{2}}{3} : 1.
	\end{gathered}
\end{equation}

For the coupling to the $\rho$ meson we adopt
\begin{equation}
		g_{\rho \Lambda} : g_{\rho\Sigma} : g_{\rho \Xi} : g_{\rho N} = 0 : 1 : 1: 1.
\end{equation}
In addition, $g_{\phi N} = 0$  is imposed on the basis that nucleons contain no strange valence quarks while the $\phi$ meson is composed of $s\bar s$. The scalar couplings  $g_{\sigma b}$ are fixed by reproducing hyperon potentials in symmetric nuclear matter as extracted from hypernuclear data~\cite{Hashimoto:2006aw,Gal:2016boi}. In particular we adopt the ratios $g_{\sigma\Lambda}/g_{\sigma N} = 0.620$, $g_{\sigma\Sigma}/g_{\sigma N} = 0.461$ and $g_{\sigma\Xi}/g_{\sigma N} =0.310$.

\begin{table}[htpb]
	\centering
	\begin{tabular}{c c c}
		\hline
		\textbf{Models}
        & 
        \textbf{NL3}$\boldsymbol{\omega\rho}$~\cite{Lalazissis:1996rd, Carriere:2002bx}
        & 
		\textbf{FSU2H}~\cite{Tolos:2016hhl,Tolos:2017lgv}\\
		\hline
		$m_\sigma$ (MeV) 
		& $497.479$ 
		& $497.479$ \\
		
		$m_\omega$ (MeV) 
		& $782.500$ 
		& $782.500$ \\
		
	    $m_\rho$ (MeV) 
		& $763.000$
		& $763.000$ \\

		$g^2_{\sigma N}$
		& $104.387$ 
		&$102.7200$ \\
		
	    $g^2_{\omega N}$
		& $165.585$ 
		&$169.5315$ \\
		
		$g^2_{\rho N}$
		& $127.000$ 
		& $197.2692$\\
		
		  $\kappa$
		& 3.860 
		& $4.0014$\\
		
		  $\lambda$
		& $-0.016$ 
		& $-0.013298$ \\
		
		$\xi$
		& $0.000$ 
		& $0.008$ \\
		
	    $\Lambda_\omega$
		& $0.030$ 
		& $0.045$ \\
	
		\hline
	    $\rho_0$ (fm$^{-3}$)
		& $0.148$ 
		& $0.1505$ \\
		
		$ \mathcal{E}_B$ (MeV)
		& $-16.3$ 
		& $-16.28$\\

		$K$ (MeV)
		& $271.6$ 
		& $238.0$ \\
		
		$m^*_N/m_N$
		& $0.60$ 
		& $0.593$ \\
		
		$E_{\rm sym}$($\rho_0$) (MeV)
		& $31.7$ 
		& $30.5$ \\
		
		$L$ (MeV)
		& $55.5$ 
		& $45.5$\\
		\hline
	\end{tabular}
	\caption{Parameters of the  NL3$\omega\rho$ and FSU2H models. Also reported are
		the values in nuclear matter at the saturation density ($\rho_0$) for
		the energy per particle ($ \mathcal{E}_B$), compressibility ($K$), effective nucleon mass ($m^*
		_N /m_N$) in symmetric nuclear matter, symmetry energy ($E_{\rm sym}$), and slope parameter of the symmetry energy ($L$).}
	\label{tab.Params_nlw_fus2h}
\end{table}


\end{document}